\documentclass[fleqn,usenatbib]{mnras}
\usepackage{newtxtext,newtxmath}

\usepackage{graphicx}	
\usepackage{amsmath}	
\usepackage{amssymb}	
\usepackage{tabularx}
\usepackage{subfigure}
\usepackage[flushleft]{threeparttable}
\usepackage{hyperref}

\newcommand{\Msun}{~M_\odot}

\newcommand{\ergs}{\rm ~erg~s^{-1}}
\newcommand{\ergcms}{\rm ~erg~cm^{-2}~s^{-1}}

\title[$\gamma$-ray emission from supernova remnants interacting with molecular clouds]{Gamma-ray emission from middle-aged supernova remnants interacting with molecular clouds: the challenge for current models}

\author[Xiaping Tang]{
Xiaping Tang,$^{1,2}$\thanks{E-mail:  tangxiaping@gmail.com, tang.xiaping@mail.huji.ac.il}
\\
$^{1}$Max Planck Institute for Astrophysics,
Karl-Schwarzschild-Str. 1,
D-85741 Garching, Germany\\
$^{2}$ The Racah Institute of physics, The Hebrew University of Jerusalem, Jerusalem 91904, Israel\\\
}

\date{Accepted XXX. Received YYY; in original form ZZZ}

\pubyear{2017}

\begin{document}
\label{firstpage}
\pagerange{\pageref{firstpage}--\pageref{lastpage}}
\maketitle

\begin{abstract}
We compare the $\gamma$-ray spectra from 10 middle-aged supernova remnants (SNRs), which are interacting with molecular clouds (MCs), with the model prediction from widely used escaping scenario and direct interaction scenario. It is found that the $\gamma$-ray data is inconsistent with the escaping scenario statistically, as it predicts a diversity of spectral shape which is not observed. The inconsistency suggests that the free escape boundary adopted in the escaping model is not a good approximation, which challenges our understanding of cosmic ray (CR) escaping in SNRs. In addition, we show that ambient CRs is potentially important for the $\gamma$-ray emission of illuminated MCs external to W28 and W44. In direct interaction scenario, the model involving re-acceleration of pre-existing CRs and adiabatic compression is able to explain the emission from most SNRs. The dispersion shown in the TeV data is naturally explained by different acceleration time of CR particles in SNRs. Re-acceleration of pre-existing CRs suggests a transition of seed particles, which is from thermal injected seed particle in young SNRs to ambient CRs in old SNRs. The transition needs to be tested by future multi-wavelength observation. In the end, we propose that radiative SNR without MC interaction is also able to produce a significant amount of $\gamma$-ray emission. A good candidate is S147. With accumulated Fermi data and CTA in future we expect to detect more remnants like S147. 


\end{abstract}


\section{Introduction}
In the past few years, both space-based GeV observatories ({\it Fermi} and {\it AGILE}) and ground-based TeV observatories ({\it H.E.S.S, MAGIC} and {\it VERITAS}) detect $\gamma$-ray emission from several middle aged SNRs, e.g. W44 \citep{Abdo10a,Giuliani11}, IC443 \citep{Albert07,Acciari09,Abdo10b}, W28 \citep{Aharonian08,Abdo10c,Hanabata14} and W51C \citep{Abdo09,Aleksic12,JF16}. Multi-wavelength observations further reveal the spatial correlation between the $\gamma$-ray emission region and the MC interaction region with robust tracers like OH maser and/or powerful diagnostic like molecular line broadening, see e.g. \cite{Jiang10} and \cite{Slane15}. 
The $\gamma$-ray emission is produced either by energetic electrons with Bremsstrahlung and Inverse Compton (IC) emission mechanism or accelerated protons with $\pi^0$-decay emission mechanism. Since dense MCs are ideal sites for proton-proton interaction, the observed spatial association with MCs implies the $\gamma$-ray emission is likely to have a hadronic origin. The characteristic $\pi^0$-decay signature around $67.5$MeV is considered to be a unique feature to distinguish hadronic emission from leptonic emission unambiguously. Recently, the $\pi^0$-decay signature is proposed in SNRs W44, IC443 \citep{Ackermann13} and W51C \citep{JF16}, which is believed to be the first direct evidence for CR proton acceleration in SNRs, making old SNRs interacting with MCs (hereafter SNR/MC) an important class of objects in $\gamma$-ray sky.   

Despite above exciting progress in observation, our theoretical understanding about CR acceleration and emission in old SNRs are still very limited. It is partly because the evolution of old SNRs is more complicated and is strongly affected by the surrounding interstellar medium. To date, two scenarios are developed to explain the MC association and the $\gamma$-ray emission with hadronic origin. One is the direct interaction scenario \citep{bykov00,Uchiyama10,TC14,TC15,Lee15, Cardillo16}, in which the remnant directly interacts with the MCs. The collision creates a cooling shock region with enhanced density and magnetic fields, where the accelerated protons and electrons are able to produce enhanced $\pi^0$-decay emission in $\gamma$-ray and synchrotron emission in radio respectively. The other one is the escaping scenario \citep{AA96,Gabici09,Fujita09,LC10,Ohira11}, in which the MCs interact with the CR particles escaping from an adjacent SNR passively. Due to the high density and magnetic fields in the MCs, runaway CR protons and electrons are able to illuminate the clouds in $\gamma$-ray with $\pi^0$-decay emission and in radio with synchrotron emission respectively. In the following discussion, the non-thermal particles in the vicinity of a SNR are referred to as CR particles while the pre-existing CR background is instead referred to as ambient CRs.


The growing number of SNR/MC detected in $\gamma$-ray enable us to investigate their physical properties in general and put better constraint on the theoretical models. Previous studies about $\gamma$-ray emission in SNR/MC focus on the individual source. In this work, we compare the $\gamma$-ray spectra of 10 SNR/MC in the First {\it Fermi} SNR Catalog \citep{Acero16} to obtain deeper insight into the physical origin of the $\gamma$-ray emission. As a first attempt, in this study, we mainly focus on the shape of $\gamma$-ray spectra without providing detailed modeling for each individual SNR/MC. 

In section \ref{sec:spectrum}, we present the $\gamma$-ray spectra from a sample of 10 SNR/MC and then discuss the interesting features shown in the spectra. In section \ref{sec:escape} and \ref{sec:interaction}, we discuss the escaping scenario and the direct interaction scenario in detail respectively. We start with a brief introduction to the scenario and then compare the model spectrum with observation to gain more insight into the origin of $\gamma$-ray emission.  Section \ref{sec:discussion} is the discussion section.

\section{Comparison of $\gamma$-ray spectrum}\label{sec:spectrum}
Based on the spatial overlap of {\it Fermi} detection and the radio extension of known SNRs, {\it Fermi} collaboration classified 30 sources as likely GeV SNRs in their First Supernova Remnant Catalog \citep{Acero16}. According to multi-wavelength observation, 11 of the 30 sources are further identified as SNR/MC. In this paper, we focus on 10 of the 11 SNR/MC, as HB21 in the 1st {\it Fermi} SNR catalog was not detected in the recent work by \cite{Ackermann17}.
 

\subsection{$\gamma$-ray spectrum from 10 SNR/MC}
\begin{figure}
\begin{center}
\includegraphics[width=\columnwidth]{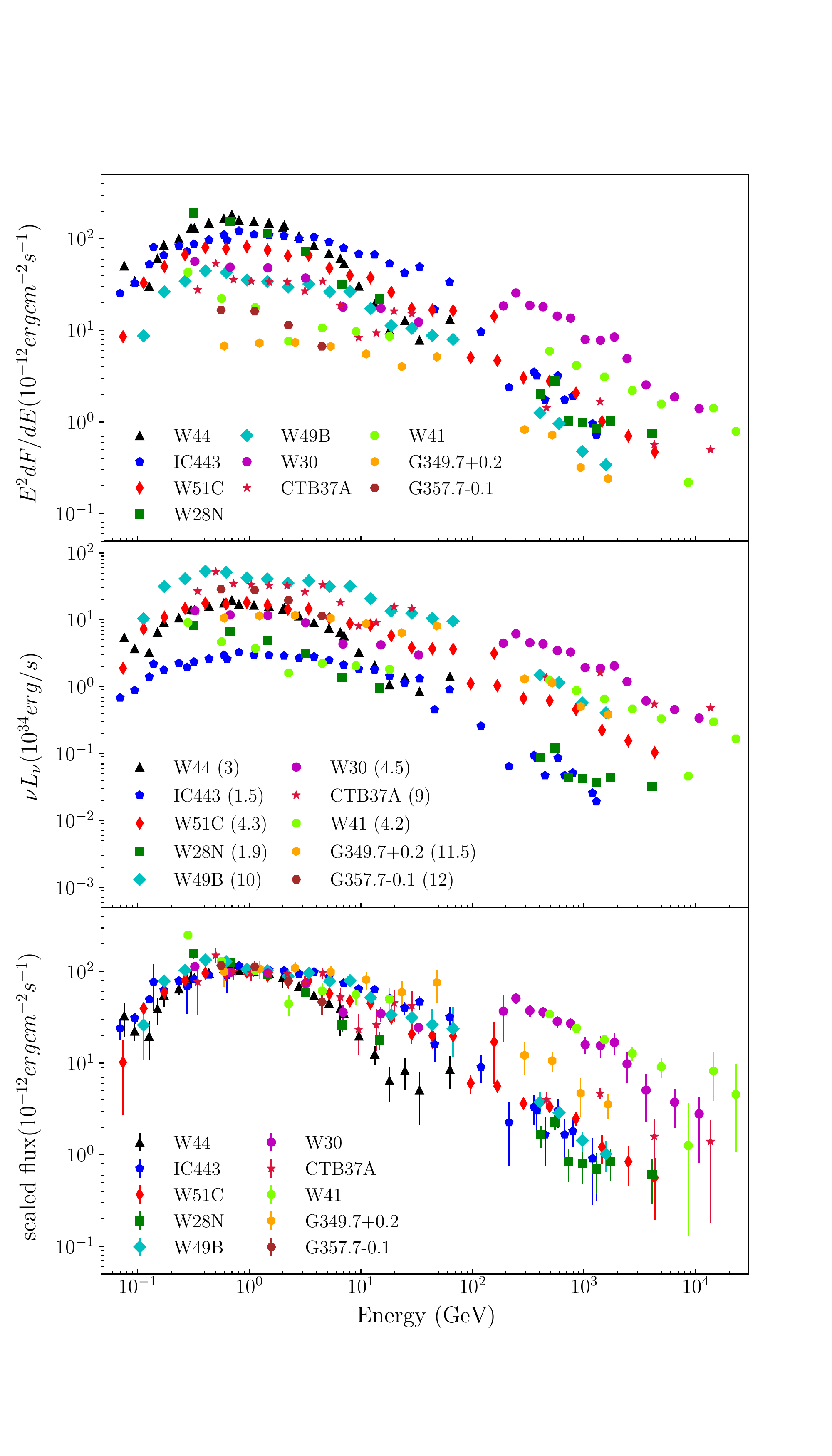} 
\caption{Upper panel: $\gamma$-ray flux $E^2dF/dE$ as a function of photon energy. Middle panel: $\gamma$-ray luminosity as a function of photon energy. Distance adopted for the luminosity calculation is indicated in the brackets in kpc. Lower panel: scaled $\gamma$-ray flux as a function of photon energy, where we normalize the flux around 1GeV to be about $10^{-10}ergcm^{-2}s^{-1}$. Reference: W44 and IC443 \citep{Ackermann13}, W51C \citep{JF16}, W28N \citep{Abdo10c}, W49B \citep{HESS16}, W30 \citep{Ajello12}, CTB 37A \citep{Aharonian08b,Brant13}, W41 \citep{Castro13,HESS15b}, G349.7+0.2 \citep{HESS15a}, G357.7-0.1 \citep{Castro13}.} 
    \label{spectrum}
\end{center}
\end{figure}

{\label{sec:spectral_property}}
In Fig. \ref{spectrum}, we present the $\gamma$-ray flux (upper panel) and luminosity (middle panel) of 10 SNR/MC with data available in the literature. The distances of all SNR/MC are taken from Table 6 in \cite{Acero16}. For W28, only the emission from the northern part of the remnant, which is spatially overlapped with HESS J1801--233, is plotted. In the lower panel, we scale the $\gamma$-ray flux around 1GeV to be $\sim 10^{-10}\ergs cm^{-2}$ to compare the shape of all 10 SNR/MC spectra. Statistical error bars are also included in the lower panel to demonstrate the comparison more clearly. 

According to Fig. \ref{spectrum}, most of the $\gamma$-ray spectra peak at a few GeV. Below $\sim 1$GeV, the spectra is characterized by a rising feature, which is consistent with the $\pi^0$-decay signature. Above $\sim 1$GeV, the spectra start to steepen and follow roughly a power law profile with no clear sign for an exponential like cutoff. The GeV emission data from {\it Fermi} and {\it AGILE} appear to be smoothly connected with the TeV emission data from {\it H.E.S.S, MAGIC} and {\it VERITAS} except W30 and W41, which implies that the GeV and TeV emission are probably generated at the same region with the same emission mechanism. 

In W30\footnote{In the GeV band, only Fermi data from the source E is taken into account \citep{Ajello12}.} (G8.7-0.1) and W41\footnote{H.E.S.S data presented for W41 is a combination of the emission from the central region and the angular region \citep{HESS15b}.} (G23.3-0.3), the TeV emission is much harder than the GeV emission and is spatially overlapped with a PWN in the vicinity of SNR/MC \citep{Ajello12,HESS15b}. It is possible that the excess of TeV emission is due to a PWN in the line of sight. Multi-wavelength observation of the two remnants is needed in future to distinguish the contribution from the PWN. In addition, the low energy spectrum of W41 doesn't show the rising feature, which might be because the low energy emission from W41 is not dominated by $\pi^0$-decay.

  
G357.7-0.1 and W44 are not detected in the TeV band. The GeV flux of G357.7-0.1 is lower than the majority of SNR/MC. Hence, the TeV emission of G357.7-0.1 may have similar slope as the other SNR/MC, but is still below the detection limit. W44 is a well-known SNR/MC and is very bright in the GeV band. The non-detection of TeV emission in W44 indicates that its TeV spectrum is much softer than the other SNR/MC. In section \ref{sec:challenge2}, we provide a possible explanation to the lack of TeV detection in W44.

The $\gamma$-ray luminosity of all 10 SNR/MC at $1$GeV varies from $10^{34}\ergs$ to $10^{36}\ergs$. The diversity in $\gamma$-ray luminosity is probably the result of different MC environments or/and different physical properties of SNRs. 

To explore the origin of observed hadronic like emission, we perform maximum likelihood fittings of the $\gamma$-ray data with both a power law (PL) and a smoothly broken power law (BPL) proton spectrum in the momentum space. The BPL proton spectrum as a function of momentum is assumed to be 
\begin{equation}
\frac{dN_p}{dp}\propto p^{-\alpha_1}\left[1+\left(\frac{p}{p_{br}} \right)^{(\alpha_2-\alpha_1)/w} \right]^{-w},
\end{equation}
where $p_{br}$ is the break momentum, $\alpha_1$ and $\alpha_2$ are power law index below and above the break respectively. $w$ determines the smoothness of the break and is fixed at 0.1 as in \cite{Ackermann13}. 

It is found that 5 SNR/MC prefer a BPL proton spectrum, which are presented in Table \ref{table:BPL}. Moreover, all 5 SNR/MC have similar $\alpha_1\sim 2.3$, $\alpha_2\sim 3.0$ and $p_{br}\sim 150$GeV/c except W44, which exhibits a much steeper spectrum with no TeV detection. Recently, \cite{Neronov17} found that $\gamma$-ray emission from nearby giant MCs in the Gould Belt also prefers a BPL proton spectrum with $\alpha_1= 2.33\substack{+0.06\\-0.08}, \,\alpha_2=2.92\substack{+0.07\\-0.04}$ and $p_{br}=18.35\substack{+6.48\\-3.57}$. It is interesting that the BPL proton spectrum derived from SNR/MC here has similar $\alpha_1$ and $\alpha_2$ but larger $p_{br}$ compared to that obtained in the isolated giant MCs. Since the $\gamma$-ray emission from isolated giant MCs traces the ambient CRs, the trend discussed above is likely an indication for re-acceleration of pre-existing ambient CRs in SNRs and will be discussed in section \ref{sec:RPCR} with details. 

The fitting results shown in Table \ref{table:BPL} are derived with the $naima$ python package \citep{Zabalza15} and only statistical errors are taken into account in the calculation. We got slightly different results for W44 and IC 443 comparing with \cite{Ackermann13}. It is possibly because the $naima$ package applies the new parameterization of the $\pi^0$-decay cross sections in \cite{Kafexhiu14}, while \cite{Ackermann13} adopt the parameterized cross section in \cite{Kamae06}. \cite{Kafexhiu14} pointed out that the parameterization provided in \cite{Kamae06} show some unphysical features around the threshold energy. Besides, the formulae developed in \cite{Kamae06} do not agree with the recent measurements of pp inelastic cross section by the TOTEM collaboration at the Large Hadron Collider \citep[e.g.,][]{Antchev13}.

The rest 5 SNR/MC need updated data in future to test whether a BPL proton spectrum is preferred to explain the $\gamma$-ray emission. As G357.7-0.1 has only 4 data points in the GeV band while the TeV emission of W30 and W41 is likely contaminated by a PWN in the line of sight. CTB 37A and W28N lack data points below $1$GeV that are important to constrain the low energy part of CR proton spectrum.

\begin{table}
\caption{$\pi^0$-decay fitting results for 5 SNR/MC with broken power law proton spectrum }
\begin{threeparttable}
\begin{tabular}{ccccc}
\hline\hline
object& $\alpha_1$&$\alpha_2$& $p_{br}$ (GeV/c) &Ref\\
\hline
IC443 &$2.28\pm 0.02$ &$3.27\pm 0.1$&$178\substack{+39\\-32}$  & 1\\
W44 &$2.29\pm 0.06$ &$3.74 \substack{+0.17\\-0.15} $ & $33\substack{+6\\-5} $& 1 \\
W51C &$2.39\pm 0.03$ &$2.91 \pm 0.06$ &$112\substack{+32\\-35}$ & 2 \\
W49B &$2.31\pm 0.03$ &$3.0\substack{+0.09\\-0.06}$ &$135 \substack{+68\\-32}$& 3\\
G349.7+0.2&$2.21\substack{+0.2\\-0.15}$ &$2.74\pm 0.14$ &$180 \substack{+130\\-80}$ & 4 \\
\hline
\end{tabular}
\label{table:BPL}
\begin{tablenotes}
\small
\item  From left to right, it is the object name, power law index below the break, index above the break, break momentum, and the reference for data used in the fitting. The fitting parameters are calculated with the Naima Python package \citep{Zabalza15} and only statistical errors are taken into account in the calculation. Reference: 1. \cite{Ackermann13}, 2. \cite{JF16}, 3. \cite{HESS16}, 4. \cite{HESS15a}. 
\end{tablenotes}
\end{threeparttable}
\end{table}

\subsection{$\pi^0$-decay signature}{\label{sec:pion_decay}}
In the literature, the rising feature shown in the ${\rm log}(E^2dF/dE)-{\rm log}(E)$ plot around a few hundreds MeV is often referred to as the $\pi^0$-decay signature and considered to be the unique feature to identify hadronic emission in SNRs. However, as we will show later, the physical interpretation of this rising feature is not trivial, which require further clarification as pointed out by \cite{Strong16}. 

In the proton-proton interaction, the dominant channel for $\gamma$-ray production is through the decay of secondary $\pi^0$, i.e. proton-proton collision creates $\pi^0$, which then quickly decays into two photons. If we assume isotropic decay of $\pi^0$ in its rest frame, then the observed $\gamma$-ray emission is found to be symmetric about half of the $\pi^0$ mass (67.5 MeV) in the ${\rm log}(dF/dE)-{\rm log}(E)$ plot \citep{Stecker71}. More importantly, above symmetry in the $\pi^0$-decay emission is independent of the primary proton spectrum, which becomes a unique feature to identify hadronic emission and is usually referred to as the $\pi^0$-decay signature. 

\begin{figure}
\begin{center}
\includegraphics[width=\columnwidth]{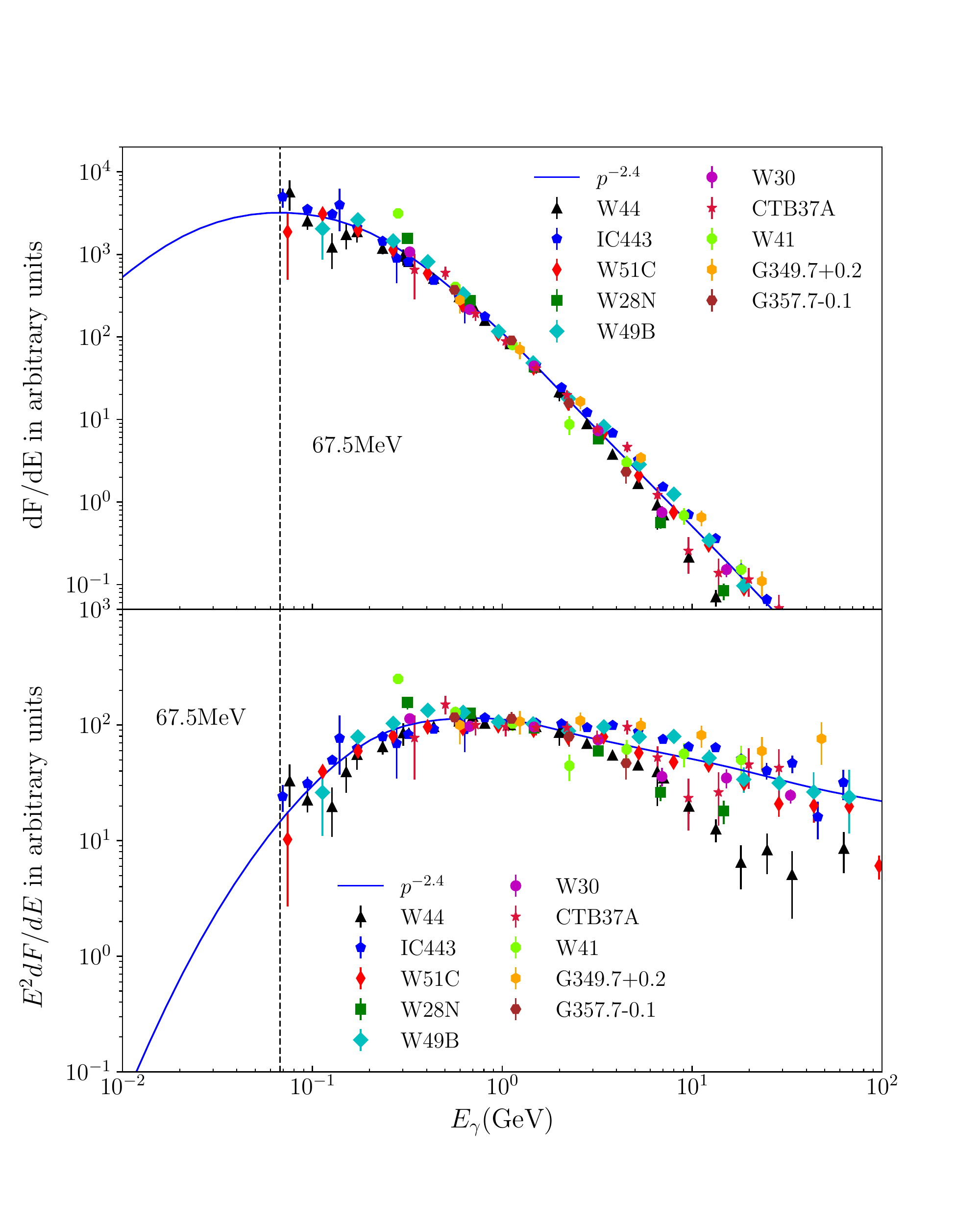} 
\caption{Upper panel: $\gamma$-ray flux $dF/dE$ as a function of photon energy $E_\gamma$ in arbitrary units. Lower panel: $\gamma$-ray flux $E^2dF/dE$ as a function of photon energy $E_\gamma$ in arbitrary units. The color points are the scaled $\gamma$-ray spectra taken from the lower panel of Fig. \ref{spectrum}. The blue solid line indicates $\pi^0$-decay emission for a primary proton spectrum $n(p)\propto p^{-2.4}$, where $p$ is the proton momentum. The black dashed line shows the photon energy of 67.5MeV. } 
    \label{fig:pion_decay}
\end{center}
\end{figure}

In the context of SNR/MC, the $\pi^0$-decay emission is produced by the interaction between primary CR protons and thermal nuclei in the vicinity of SNRs. In the upper panel of Fig. \ref{fig:pion_decay}, we show the $\pi^0$-decay emission for a primary proton spectrum $n(p)\propto p^{-2.4}$ (blue solid line) in ${\rm log}(dF/dE)-{\rm log}(E)$ plot, where $p$ is the proton momentum. The $\pi^0$-decay signature is clearly illustrated in the plot as a symmetric bump like feature around $67.5$MeV (dashed line). $\gamma$-ray data below 67.5 MeV is essential to reveal the symmetric $\pi^0$-decay signature unambiguously.

In the widely used ${\rm log}(E^2dF/dE)-{\rm log}(E)$ plot, due to the multiplication of $E^2$ factor, the symmetric bump like feature changes into a rising feature with a flattening of spectral slope around $100$MeV, see the lower panel of Fig. \ref{fig:pion_decay}. Since the unique symmetric feature of $\pi^0$-decay is now hidden in the rising feature and reflected as a changing of spectral slope, the identification of $\pi^0$-decay signature becomes much more difficult in the ${\rm log}(E^2dF/dE)-{\rm log}(E)$ plot. Here, we want to emphasize that it is more straightforward to search for $\pi^0$-decay signature in the ${\rm log}(dF/dE)-{\rm log}(E)$ plot than the commonly used ${\rm log}(E^2dF/dE)-{\rm log}(E)$ plot. 

In Fig. \ref{fig:pion_decay}, we also present the scaled $\gamma$-ray spectra from all 10 SNR/MC. In the upper panel, it is found that below $\sim$100MeV the scaled $\gamma$-ray spectra bend toward a harder spectrum with decreasing energy, which is consistent with the symmetric $\pi^0$-decay signature. However, the lack of data points below 67.5 MeV prevents us from identifying the $\pi^0$-decay signature unambiguously, as the above bending feature can also be explained by Bremsstrahlung emission with a BPL electron spectrum. In the lower panel, the scaled $\gamma$-ray spectra are consistent with a rising feature. But it is less clear why the $\gamma$-ray data below 67.5 MeV is crucial for identifying the $\pi^0$-decay signature. This again demonstrates that $\pi^0$-decay signature is better revealed in the ${\rm log}(dF/dE)-{\rm log}(E)$ plot instead of ${\rm log}(E^2dF/dE)-{\rm log}(E)$ plot.

\subsection{Hadronic origin of $\gamma$-ray emission}
Next, we want to discuss how to identify hadronic emission in SNR/MC without data below $\sim 100$MeV, as both {\it Fermi} and {\it AGILE} don't have sensitivity below $\sim 100$MeV. Recently, hadronic emission is identified in W44, IC443  and W51C \citep{Giuliani11,Ackermann13,JF16} based on a combination of several factors. The first one is the detection of a rising feature in $(E^2dF/dE)-{\rm log}(E)$ plot which is consistent with $\pi^0$-decay. Secondly, the spatial correlation between the $\gamma$-ray emission region and the MC interaction region also favors hadronic origin of $\gamma$-ray emission \citep[e.g.,][]{Jiang10,Slane15}. In the end, detailed calculations show that both IC and Bremsstrahlung emission mechanism can not reproduce the $\gamma$-ray emission naturally. With typical background photon field, the simulated IC emission is too low to explain the $\gamma$-ray data. Bremsstrahlung emission is able to reproduce the rising feature in observation but requires an internal break in the electron spectrum. However, there is no physical reason for a break around a few hundreds MeV in the electron spectrum. Besides, Bremsstrahlung emission can not explain the TeV emission from SNR/MC. It is the combination of all these factors together which makes us believe that the $\gamma$-ray emission from W44, IC443  and W51C have a hadronic origin.


Based on above discussion, we constrain our study to hadronic models in the rest of this paper. IC and Bremsstrahlung emission is assumed to be negligible for our discussion, which should be a good approximation for energy $\gtrsim1$GeV.  With hadronic origin, the rising feature below $\sim 1$ GeV in the observed $\gamma$-ray spectra of SNR/MC can be naturally explained by the $\pi^0$-decay signature. Although the ${\rm log}(E^2dF/dE)-{\rm log}(E)$ plot can not reveal the $\pi^0$-decay signature very clearly, it provides more details about the spectral shape at high energy part. In the following sections, we mainly focus on the features in the high energy part of spectra above $\sim 1$ GeV. As a result, all the figures will be in the ${\rm log}(dF/dE)-{\rm log}(E)$ format.

In the next two sections, we describe the escaping scenario and direct interaction scenario in detail and then compare the model spectra with observation. The main difference between the two scenarios is the source of primary CR protons. The escaping scenario focuses on the CR particles that escaped from the remnant, while the direct interaction scenario instead investigates energetic particles confined within the remnant. In both scenarios, the CR particles are believed to be accelerated at the remnant shock through the diffusive shock acceleration (DSA) process\citep[e.g.,][]{Bell78,B&E87}. 

In current non-linear theory of DSA , there are still two open questions, one is how do energetic particles manage to escape the shock region and the other is how are seed particle injected into the DSA process. Both problems are not fully understood at this point \citep[e.g.,][]{M&D01} and require a special prescription in the treatment of DSA. In young SNRs, free escape boundary is widely used to describe the particle escaping and thermal injection of seed particles is often assumed for particle injection. In middle aged SNRs, both prescriptions, however, confront some challenges which will be discussed in the following sections.

In this paper, $\pi^0$-decay emission from the proton-proton interaction is calculated with the parameterized $\gamma$-ray production cross sections derived in \cite{Kafexhiu14}. The formula is found to be accurate within $20\%$ accuracy from the kinematic threshold ($280$MeV) up to PeV energies. At low energy, the model is fitted with experimental data while at high energy it is tested with public available code results. Please see Appendix \ref{app:pion_decay} for more details. 


\section{Escaping scenario}\label{sec:escape}
\subsection{Basic idea}
CR particles accelerated at SNR shock can escape from the remnant and then propagate into the surrounding interstellar medium (ISM) after gaining enough energy. When these escaping CR particles encounter a dense MC, they interact with thermal nuclei in the MCs and illuminate the clouds in the $\gamma$-ray sky through $\pi^0$-decay emission \citep{AA96}. Assuming the remnant is a point source, \cite{Gabici09} modeled the multi-wavelength emission of an illuminated MC in detail. Later, \cite{LC10} and \cite{Ohira11} extended the model to account for the finite size of a SNR.

Pre-existing ambient CRs in the ISM are also able to interact with particles in MCs and produce a significant amount of $\gamma$-ray emission. If we extrapolate the $\gamma$-ray emission detected in nearby giant MCs to arbitrary distance $d$, the $\gamma$-ray contribution from the interaction between ambient CRs  and giant MCs at $\sim 3$GeV  is approximate \citep{Yang14}
\begin{equation}
F_{am}\sim 2\times 10^{-11}\ergcms \left(\frac{M}{10^5\Msun}\right)\left(\frac{d}{1kpc}\right)^{-2},
\label{eq:ambient_CR}
\end{equation}
which is unimportant for most of SNR/MC discussed here.

\subsection{Runaway CR spectrum and the $\pi^0$-decay emission from illuminated MCs}\label{sec:runaway_spectrum}
\begin{figure}
\begin{center}
\includegraphics[width=\columnwidth]{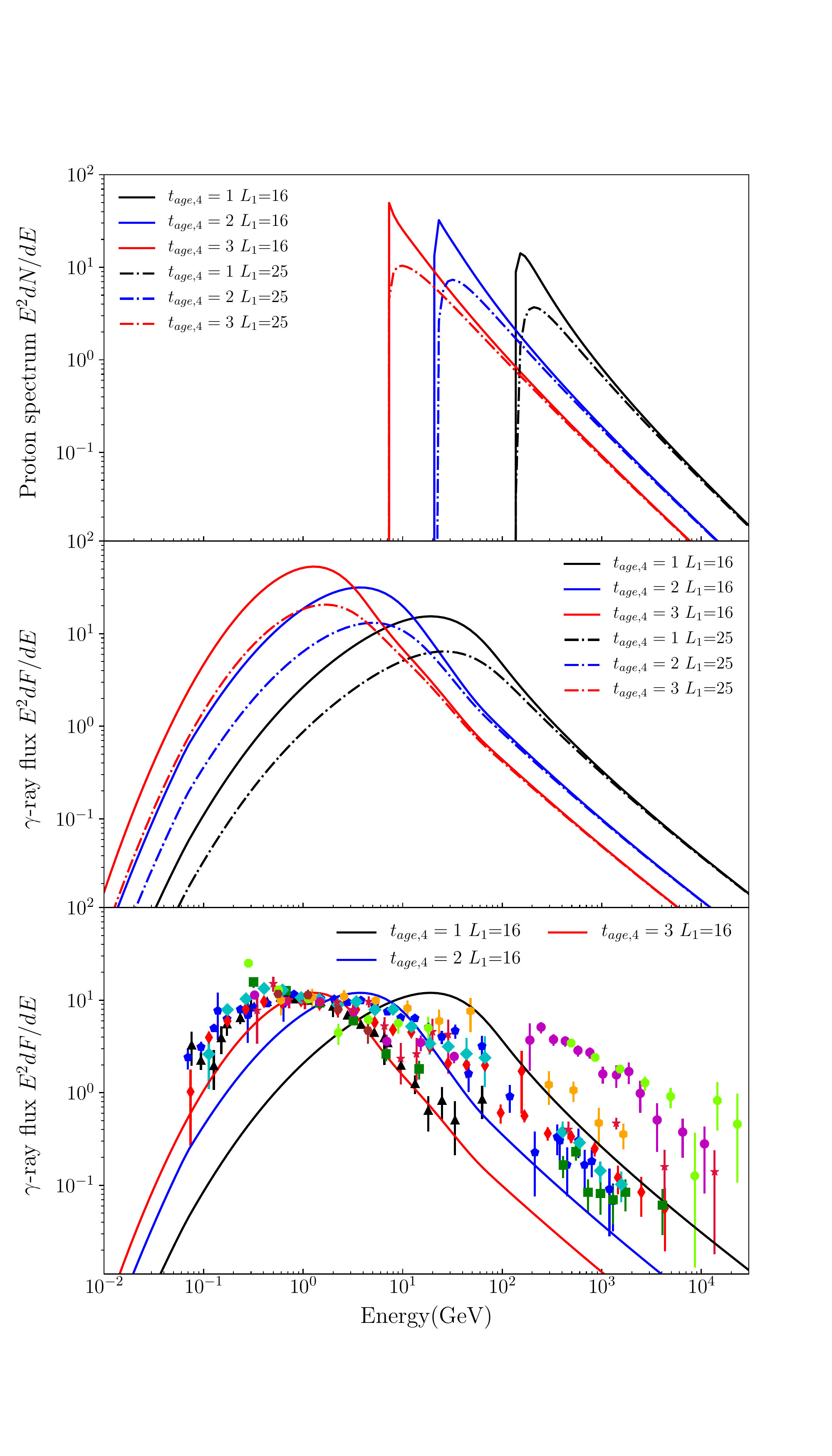} 
\caption{From upper to lower panel, it is the runaway CR proton spectrum $E^2dN/dE$ as a function of proton kinetic energy, the corresponding $\pi^0$-decay emission $E^2dF/dE$ as a function of photon energy, and the scaled $\gamma$-ray data and model spectra as a function of photon energy. $t_{age,4}$ is the remnant age in $10^4$yrs and $L_1$ is the distance between the inner radius of MCs and the center of the SNR in pc. See Appendix \ref{app:runaway_spectrum} for more details about the calculation.} 
    \label{fig:escape}
\end{center}
\end{figure}

In this section, we compare the $\gamma$-ray emission predicted from escaping model with that measured in observation. In the upper panel of Fig. \ref{fig:escape}, we present the runaway CR proton spectrum in a illuminated MC for different $t_{age}$ and $L_1$, where $t_{age}$ is the age of SNR and $L_1$ is the distance between the SNR and the nearby MCs. The runaway CR spectrum is calculated with the model in \cite{Ohira11} which is described with details in Appendix \ref{app:runaway_spectrum}. Different combination of $t_{age}$ and $L_1$ corresponds to a different spatial configuration of SNR/MC, which are used to demonstrate the diversity of $\gamma$-ray spectra expected from the escaping scenario. In the lower panel of Fig. \ref{fig:escape}, we compare the model spectra with $\gamma$-ray data. Given the complexity and diversity of all SNR/MC, we focus on only the comparison of spectral shape, where the normalization factor is left as a free parameter. Both the model spectra and $\gamma$-ray data are scaled to have the same peak value.

According to Fig. \ref{fig:escape}, the runaway CR proton spectrum is characterized by a sharp low energy cutoff $E_{low}$, which is a natural result of the free escape boundary adopted in the escaping scenario. Under the assumption of free escape boundary, CR particles with higher energy escape from the remnant earlier during the SNR evolution and there is a threshold energy $E_{low}$ for escaping CRs.  Particles with energy $E<E_{low}$ either haven't escaped from the remnant yet or do not have enough time to diffuse into the nearby MCs, thus leaving a low energy cutoff $E_{low}$ in the runaway CR spectrum. The sharp cutoff feature is not shown in the corresponding $\pi^0$-decay emission. It is mainly because photons emitted in $\pi^0$-decay spread in a large energy range, which smooth the sharp feature. However, the peak of $\gamma$-ray flux in escaping model does depend on $E_{low}$ and is shifted to higher energy with larger $E_{low}$. The trend is valid as long as the cutoff energy $E_{low}$ is larger than the threshold energy $280$MeV. In order to reproduce the observed $\gamma$-ray spectra, the low energy cutoff of runaway CR proton spectrum has to satisfy $E_{low}\lesssim 1$ GeV as shown in the lower panel of Fig. \ref{fig:escape}. 

\subsection{Escape of GeV CR particles}\label{sec:GeVescape}
$E_{low}\lesssim 1$ GeV brings a fundamental problem to the escaping scenario, which is can/how does the low energy CR particle ($\sim 1$GeV) escape from middle aged SNRs. Early work by \cite{Bell78} proposed that damping mechanism such as neutral-ion damping can suppress CR induced turbulence and facilitate the escaping of CR particles. Recently, \cite{PZ03} further investigate how the maximum momentum $p_{max}$ of CR particles evolves in the Sedov-Taylor (ST) phase of a SNR. 
It is argued that both non-linear wave damping and linear neutral-ion damping are crucial for the escape of low energy CR particles.

Before we continue our discussion, we want to clarify that the maximum momentum at the shock front $p_{max}$ is equivalent to the escaping momentum of the remnant $p_{esc}$ under the free escape boundary condition, i.e. particles with momentum $p>p_{esc}=p_{max}$ are able to escape the remnant through the free escape boundary. The maximum momentum of particle within the SNRs $p_{SNR}$, however, can be larger than the maximum momentum at the shock front $p_{max}$, if there are energetic CR particles trapped in the interior of a remnant. In this paper, we assume $p_{SNR}=p_{max}=p_{esc}$ for simplification.

In case of pure non-linear wave damping with a Kolmogorov-type energy cascade, the maximum momentum $p_{max}$ in the damping dominated regime with slow remnant shock satisfies \citep{PZ03}\footnote{The coefficient 0.24 in  eq. (19) of \cite{PZ03} is not correct and needs to be replaced with 24 here.}
\begin{eqnarray}
\frac{p_{max}}{m_p c}&\approx&\frac{24\kappa a^2 C^2_{cr}(a)C^3_K\xi^2_{cr}u^7_{sh}R_{sh}}{r_{g0}V_a^4c^3} \nonumber\\
&\approx &0.72~\frac{u^7_{sh}R_{sh}}{r_{g0}V_a^4c^3}  \nonumber\\
&\approx & 2\times 10^3~n_{H,0}^2B_{a,0}^{-3}u_{sh,2}^7R_{sh,1}.
\label{eq:non-linear}
\end{eqnarray}
$m_p$ is proton mass, $n_{H,0}$\footnote{If two phase ISM is considered, then $n_{H,0}$ represents the number density in the intercloud medium. } is the ambient density of hydrogen atom in $cm^{-3}$, $B_{a,0}$ is the ambient magnetic field in $\mu G$, $u_{sh,2}$ is the shock velocity in 100km/s and $R_{sh,1}$ is the remnant radius in 10pc.

In case of pure neutral-ion damping, the maximum momentum $p_{max}$ in the damping dominated regime with slow remnant shock instead becomes \citep{PZ03}
\begin{eqnarray}
\frac{p_{max}}{m_p c}&\approx & \frac{2aC_{cr}(a)\xi_{cr}u_{sh}^3}{cV_ar_{g0}\nu_{in}} \nonumber\\
&\approx &0.25~ \frac{u_{sh}^3}{c V_a r_{g0}\nu_{in}} \nonumber\\
&\approx & 2u_{sh,2}^3T_4^{-0.4}n_n^{-1}n_i^{1/2},
\label{eq:linear}
\end{eqnarray}
where $u_{sh,2}$ is the shock velocity in 100km/s, $T_4$ is the temperature in $10^4K$, $n_n$ is the neutral particle number density in $cm^{-3}$ and $n_i$ is the ion number density in $cm^{-3}$. Neutral-ion damping is only important when the precursor becomes partially ionized. This is likely to happen at slow shock with $u_{sh}\lesssim150$km/s \citep{HM89}.

In eq. (\ref{eq:non-linear}) and (\ref{eq:linear}), from the first step to the second step we apply the same $\kappa, a$ and $\xi_{cr}$ as that in \cite{PZ03}, see the detailed definition of all the parameters there. Smaller $\kappa, a$ and $\xi_{cr}$ certainly can decrease $p_{max}$ and make it easier for GeV particles to escape. But smaller $\kappa, a$ and $\xi_{cr}$ also decrease $p_{max}$ at early time evolution of a SNR and make it even more difficult for the remnant to become a PeV accelerator.  Hence, if we assume SNRs are CR accelerators up to the knee energy $\sim 10^{15}$eV, then $\kappa, a$ and $\xi_{cr}$ can't be arbitrarily small and deviate too much from those adopted in \cite{PZ03}.


According to eq. (\ref{eq:non-linear}) and (\ref{eq:linear}), the escape of GeV particles put strong constraint on the physical properties of ISM and SNR, e.g. the shock velocity $u_{sh}$ has to satisfy $u_{sh}\lesssim 100$km/s. However, when the shock velocity slows down to $u_{sh}\lesssim 200$km/s in typical ISM environment, radiative cooling becomes important. Because of cooling loss, the remnant gradually leaves the adiabatic ST phase and eventually enters the radiative phase, which further complicates the discussion. As a result, the picture proposed in \cite{PZ03} for ST phase of SNR evolution may no longer be valid for those middle-aged SNRs. Besides, it is also unclear from observation whether strong wave damping is indeed happening in these middle-aged SNRs, which deserve further investigation.


\subsection{Challenge for the model}
The main challenge for escaping scenario is how to obtain a low energy cutoff $E_{low}\lesssim 1$GeV. \cite{Ohira11} propose that the observed SNRs and MCs are very close or even in physical contact. In this special configuration, CR particles with all energy are able to escape the remnant and then diffuse into the adjacent MCs. With above assumption, \cite{Ohira11} are able to reproduce the $\gamma$-ray emission in several SNR/MC. However, when SNR and MCs are directly interacting with each other, the validity of escaping scenario becomes an open question. In the unshocked part of MCs, the $\gamma$-ray emission is possibly interpreted by the illuminated cloud model, while in the interaction region the $\gamma$-ray emission is most likely described by the direct interaction scenario. 

In escaping scenario, the shape of model spectrum strongly depends on the spatial configuration of SNR/MC . Since the observed 10 SNR/MC are likely in different spatial configuration, hence we expect to detect a variety of $\gamma$-ray spectra with different shape, which however is not seen in current data as indicated by the lower panel of Fig. \ref{fig:escape}. The diversity of $\gamma$-ray spectra  in SNR/MC expected from escaping scenario is also discussed in \cite{Gabici09}. Someone may argue that the 10 SNR/MC discussed here is biased to middle-aged SNRs in physical contact with MCs as proposed by \cite{Ohira11}. However, the illuminated MCs external to W28 and W44 also exhibit similar spectrum with strong GeV emission and peak at a few GeV. In addition, $\gamma$-ray spectrum from middle-aged SNRs with no signature of MC interaction, e.g., SNRs S147 \citep{Katsuta12} and Cygnus Loop \citep{Katagiri11}, is also similar as those in SNR/MC, which can not be explained by the escaping scenario.  

If we attribute the $\gamma$-ray emission from all these objects to illuminated clouds, then it is quite puzzling why we didn't see any objects with $\gamma$-ray spectrum like the blue and black lines in the lower panel of Fig. \ref{fig:escape}. 
$\gamma$-ray emission from many young SNRs does show strong TeV emission, see e.g., Fig. 7 in \cite{Funk15}. But the observed emission is likely dominated by leptonic emission instead of hadronic emission.
In summary, the $\gamma$-ray spectra from 10 SNR/MC are inconsistent with the prediction from escaping scenario statistically, which need to be confirmed by future observation.  This inconsistency not only challenges the escaping scenario but also challenges our understanding of CR escaping in SNRs. It might suggest that the free escape boundary widely adopted in the modeling of CR escaping is not a good assumption \citep{Drury11}. We have to keep in mind that the introduction of free escape boundary in non-linear DSA was originally intended to achieve a self-consistent treatment of particle acceleration at SNR shock front. It is not designed to investigate the distribution and propagation of escaping CR in the ISM surrounding a SNR.

\subsection{Illuminated clouds in W28 and W44}
The best candidates for illuminated MCs are $\gamma$-ray bright MCs which are adjacent to a SNR but are not in physical contact with the remnant. The $\gamma$-ray source HESS J1800-240 external to W28 (hereafter W28 240) and the $\gamma$-ray bright regions surrounding W44 (hereafter W44 MCs) are considered to be two good candidates in this category \citep{Hanabata14,Uchiyama12}. 

The environment in W28 240 is very complicated and needs further explanation before we continue our discussion \citep{Aharonian08,Hanabata14}. In the TeV band, W28 240 is resolved into 3 individual sources, 240A, 240B and 240C. 240A is spatially coincident with two HII regions, G6.1-0.6 and G6.225-0.569, while 240B is spatially associated with the ultra-compact HII region W28A2. 240C is spatially overlapped with SNR G5.71-0.08, which is likely interacting with MCs according to the OH maser detection. Please see \cite{Hanabata14} and references therein for more details.

\begin{figure}
\begin{center}
\includegraphics[width=\columnwidth]{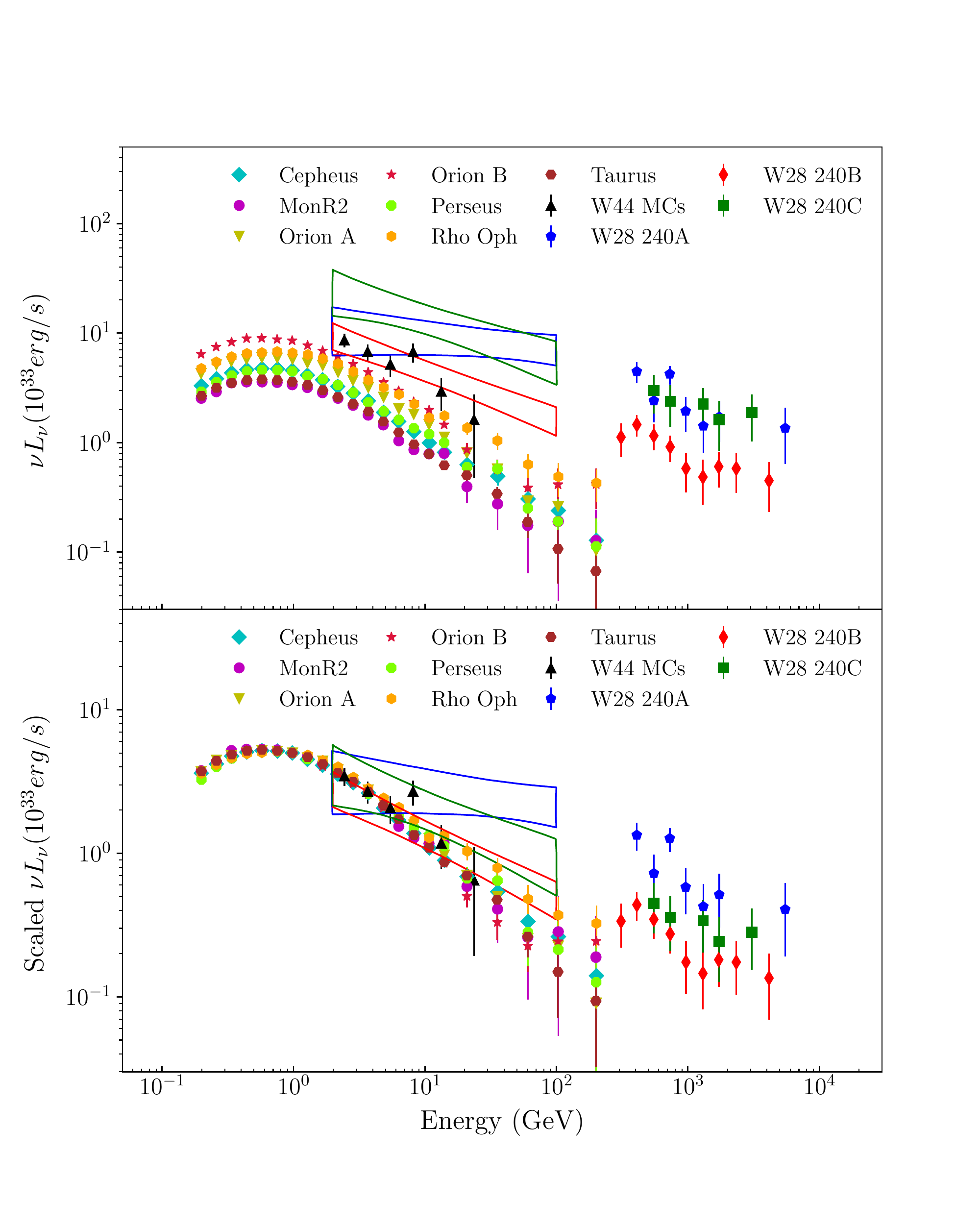} 
\caption{Upper panel: $\gamma$-ray luminosity from W44 MCs, W28 240 and 7 isolated MCs in Gould Belt. The luminosity is normalized to a cloud mass of $10^5 \Msun$ based on the mass data in Table \ref{table:GMC}. Lower Panel: scaled $\gamma$-ray luminosity to compare the spectral shape of different MCs.  Reference: isolated MCs in Gould Belt \citep{Neronov17}, W44 MCs \citep{Uchiyama12} and W28 240 \citep{Hanabata14}. The distance applied in the luminosity calculation is also listed in Table \ref{table:GMC}. }
    \label{fig:GMC}
\end{center}
\end{figure}

In this section, we compare the $\gamma$-ray emission from MCs external to W28 and W44 with emission from isolated MCs in the Gould Belt \citep{Neronov17}.
In the upper panel of Fig. \ref{fig:GMC}, we provide the $\gamma$-ray luminosity of W28 240, W44 MCs and several isolated MCs in the Gould Belt. The luminosity is normalized for a cloud mass of $10^5\Msun$ and is calculated with the distance and mass shown in Table \ref{table:GMC}. When both CO and dust estimated mass is available, the dust estimated cloud mass is applied in the calculation. It is found that the normalized $\gamma$-ray luminosity of 240B and W44 MCs is comparable with that from isolated MCs while the luminosity of 240A and 240C are higher than the isolated MCs.
Since there are CO dark clouds \citep[e.g, ][]{Grenier05}, CO emission as a tracer of MCs is likely to underestimate the cloud mass. If the cloud mass in W28 240 and W44 MCs are larger than the CO estimated value, then the corresponding normalized luminosity becomes even smaller and thus more consistent with the isolated MCs. 

In the lower panel of Fig. \ref{fig:GMC}, we present the scaled $\gamma$-ray luminosity to compare the shape of different spectra. The spectral shape of W44 MCs and W28 240B qualitatively agree with that of isolated MCs. 240C and 240A show harder spectrum, which is attributed to the escaping CRs in the literature. However, there are other possible explanations which require further investigation. 240C is spatially overlapped with SNR G5.71-0.08, which is interacting with MCs. Hence, the $\gamma$-ray emission of 240C can be explained by direct interaction scenario without involving escaping CRs. The GeV data of 240A is harder than the TeV data from H.E.S.S with a break around 100GeV. The excess of GeV emission in 240A is likely affected by the spatially coincident HII regions. Based on above discussion, the $\gamma$-ray emission in W44 MCs and W28 240 is possibly attributed to or at least partly attributed to the ambient CRs and other associated sources, which need to be tested by future observation.

\begin{table*}
\centering
\caption{Physical properties of MCs investigated in this paper.}
\begin{threeparttable}
\begin{tabular}{ccccc}
\hline\hline
object& Mass[Dust]($10^5\Msun$)& Mass[CO]($10^5\Msun$)& Distance(kpc)& Reference  \\
\hline

Cepheus&$-$&1.9&0.45& M: 1, D: 1\\
Mon R2 & $1.1$& $0.80$ & 0.83 & M: 2, D: 1\\
Orion A &  $1.2$& $0.80$ &0.5& .\\
Orion B &  $0.78$&$0.65$ &0.5& .\\
Perseus&$ 0.41$&$0.3$&0.35& .\\
Rho Oph&$0.12$&$0.08$&0.165& .\\
Taurus &$0.30$& $0.23$ & 0.14& .\\
W28 240A &$-$&$0.23$&1.9 & M: 3 and 4, D: 5\\
W28 240B &$-$&$0.7$ &1.9& .\\
W28 240C& $-$&$0.14$&1.9&. \\
W44 MCs &$-$&$5$ &2.9& M: 6 and 7, D: 6\\

\hline\hline
\end{tabular} 
\begin{tablenotes}
\small
\item From left to right, it is the object name, MC mass estimated with dust observation, MC mass estimated with CO observation, distance and reference. In the reference column, M represents the reference for mass estimation and D indicates the reference for distance estimation.  Reference: 1. \cite{Dame87}, 2. \cite{Yang14}, 3. \cite{Aharonian08}, 4. \cite{Hanabata14}, 5. \cite{Velaz02}, 6. \cite{Seta98}. 7. \cite{Uchiyama12}.  

\end{tablenotes}
\end{threeparttable}
\label{table:GMC}
\end{table*}

\section{Direct interaction scenario}\label{sec:interaction}
\subsection{Basic idea}
When a SNR collides with nearby MCs, the shock front driven by the remnant is slowed down by dense clouds. As a result, the postshock temperature drops and triggers efficient radiative cooling. Because of cooling induced compression, a thin shell with high density and magnetic field eventually forms behind the shock front. The thin radiative shell is an ideal site for both $\pi^0$-decay emission in $\gamma$-ray and synchrotron emission in radio \citep{B&C82}. The direct interaction scenario focuses on the interaction between SNRs and MCs, and is motivated by the slow MC shock detected in observation. Slow MC shocks are good indicators for direct MC interaction and have been identified in all 10 SNR/MC discussed here with robust tracers such as OH maser.

\cite{Uchiyama10} investigate the interaction between a young (non-radiative) SNR and MCs, and then adopt the model to explain the $\gamma$-ray and radio emission in W44. \cite{TC14,TC15} instead study the collision between an old (radiative) SNR and MCs. In the former case, the fast (non-radiative) shock in a young SNR is slowed down by dense MCs and then drives a radiative shell with enhanced density and magnetic field into the MCs. In this case, both the radio and $\gamma$-ray emission are expected to follow the MC interaction region in morphology. In the latter case, the old (radiative) SNR has a radiative shell itself, which can produce a significant amount of radio and $\gamma$-ray emission. The collision between SNR and MCs then creates a region with even higher density and magnetic field, which is also able to generate enhanced radio and $\gamma$-ray emission. IC 443 is considered to be a good example for this case. \cite{TC14,TC15} propose that such two components model can explain the discrepancy between radio and $\gamma$-ray morphology in SNR/MC like IC 443. 

If the picture proposed by \cite{TC14} is correct, then old (radiative) SNRs without MC interaction are also able to produce a significant amount of $\gamma$-ray emission in its cooling shell. The $\gamma$-ray luminosity of radiative SNRs without MC interaction is expected to be smaller than those SNR/MC, making it more difficult to detect such objects. S147 is probably a good example of this category with $\gamma$-ray peak luminosity of only $10^{33}\ergs$. Multi-wavelength observations of S147 show that the $\gamma$-ray, radio and H$\alpha$ emission regions are spatially correlated with each other very well \citep{Xiao08,Katsuta12}. Since $\gamma$-ray, radio and H$\alpha$ emission are expected to trace the accelerated protons, electrons and the cooling shell respectively, the spatial correlation among them is consistent with emission from a radiative SNR.  In \cite{Katsuta12}, the multi-wavelength emission from S147 is studied with the escaping scenario, but require an extremely low filling factor for the interaction region. Besides, there is no signature for MC interaction in S147. So it is more natural to attribute the $\gamma$-ray emission to the radiative shell of an old SNR without MC interaction. With accumulated Fermi data and CTA in future we expect to detect more remnants like S147. 

Next, we discuss the primary CR spectrum within the framework of direct interaction scenario. Several different ideas have been proposed to interpret the observed $\gamma$-ray emission with hadronic origin, including pure adiabatic compression of pre-existing ambient CRs, DSA of thermal injected seed particles and re-acceleration of pre-existing ambient CRs.

\subsection{Pure adiabatic compression}\label{sec:ad_compression}
We start with the simplest one, which is pure adiabatic compression of pre-existing ambient CRs in the cooling shell. The number density of ambient CR proton is  assumed to be \citep{Bisschoff16}
\begin{align}
n_{CR}(E)=\frac{4\pi J_{CR}(E)}{\beta c}\nonumber =&1.56\times 10^{-10} {\rm cm^{-3}GeV^{-1}} \nonumber \\
&\times  \frac{E_0^{1.03}}{\beta^3}\left( \frac{E_0^{1.21}+0.77^{1.21}}{1+0.77^{1.21}}\right)^{-3.18},
\label{eq:CR_spectrum}
\end{align}
where $\beta$ is the proton velocity in c, $E$ is the proton kinetic energy and $E_0=E/GeV$. The ambient CR proton spectrum provided above is able to reproduce the Voyager 1 data between 6MeV to 60 MeV and the PAMELA spectrum from 50GeV to 100GeV within about $12\%$ accuracy. 

In the radiative shell, the ambient CR spectrum is boosted by a large factor due to adiabatic compression, which then is able to produce enhanced $\pi^0$-decay emission. The CR spectrum after adiabatic compression is \citep{Uchiyama10}
\begin{equation}
n_{ad}(p)= s^{2/3}n_{CR}(s^{-1/3}p),
\label{eq:ad_spectrum}
\end{equation}
where $s$ is the adiabatic compression ratio and $n_{CR}(p)=\beta c\, n_{CR}(E)$ is the CR proton number density as a function of momentum. 

If we assume the cooling shell is supported by magnetic pressure \citep{Chevalier77,B&C82}, then we have 
\begin{equation}
\frac{B^2_{shell}}{8\pi}=\mu_H n_H u_{sh}^2,
\end{equation}
where $\mu_H$ is the mass per hydrogen nucleus, $n_H$ is the number density of hydrogen atom and $u_{sh}$ is shock velocity. Based on conservation of mass and magnetic flux, the compression ratio $s$\footnote{Note the compression ratio $s$ estimated here is $2/3$ times smaller than that presented in eq. (2) of \cite{Uchiyama10} due to a geometric factor.} in a radiative shell is estimated as \citep{Chevalier77,TC14}
\begin{equation}
s=\frac{B_{shell}}{B_a}\approx 63 \,n_{H,0}^{0.5}\, u_{sh,2} B_{a,0}^{-1},
\label{eq:compression_ratio}
\end{equation}
where $n_{H,0}$ is hydrogen number density in $cm^{-3}$, $u_{sh,2}$ is shock velocity in $100 \rm km/s$ and $B_{a,0}$ is ambient magnetic field in $\mu G$. 

When $E_0 \gg 1$, the CR spectrum provided in eq. (\ref{eq:CR_spectrum}) approaches a power law with $n_{CR}(p)\propto p^{-2.82}$. In such situation, the CR spectrum after adiabatic compression becomes 
\begin{equation}
n_{ad}(p)=s^{2/3}n_{CR}(s^{-1/3}p)\approx s^{1.6}n_{CR}(p).
\label{eq:compression_approx}
\end{equation} 
According to eq. (\ref{eq:compression_ratio}) and (\ref{eq:compression_approx}), with shock velocity $u_{sh,2}\sim 1$ and $n_{H,0}B_0^{-1}\sim 1$ in the ISM, it is easy to obtain a few hundreds of times enhancement in the CR spectrum. 
In the shell and MC interaction region discussed by \cite{TC14}, the primary CR spectrum is further boosted by a factor of few. From energetic point of view, a boost factor about a few hundreds is capable to explain the observed $\gamma$-ray luminosity in many SNR/MC \citep{TC14,Cardillo16}. 

\begin{figure}
\begin{center}
\includegraphics[width=\columnwidth]{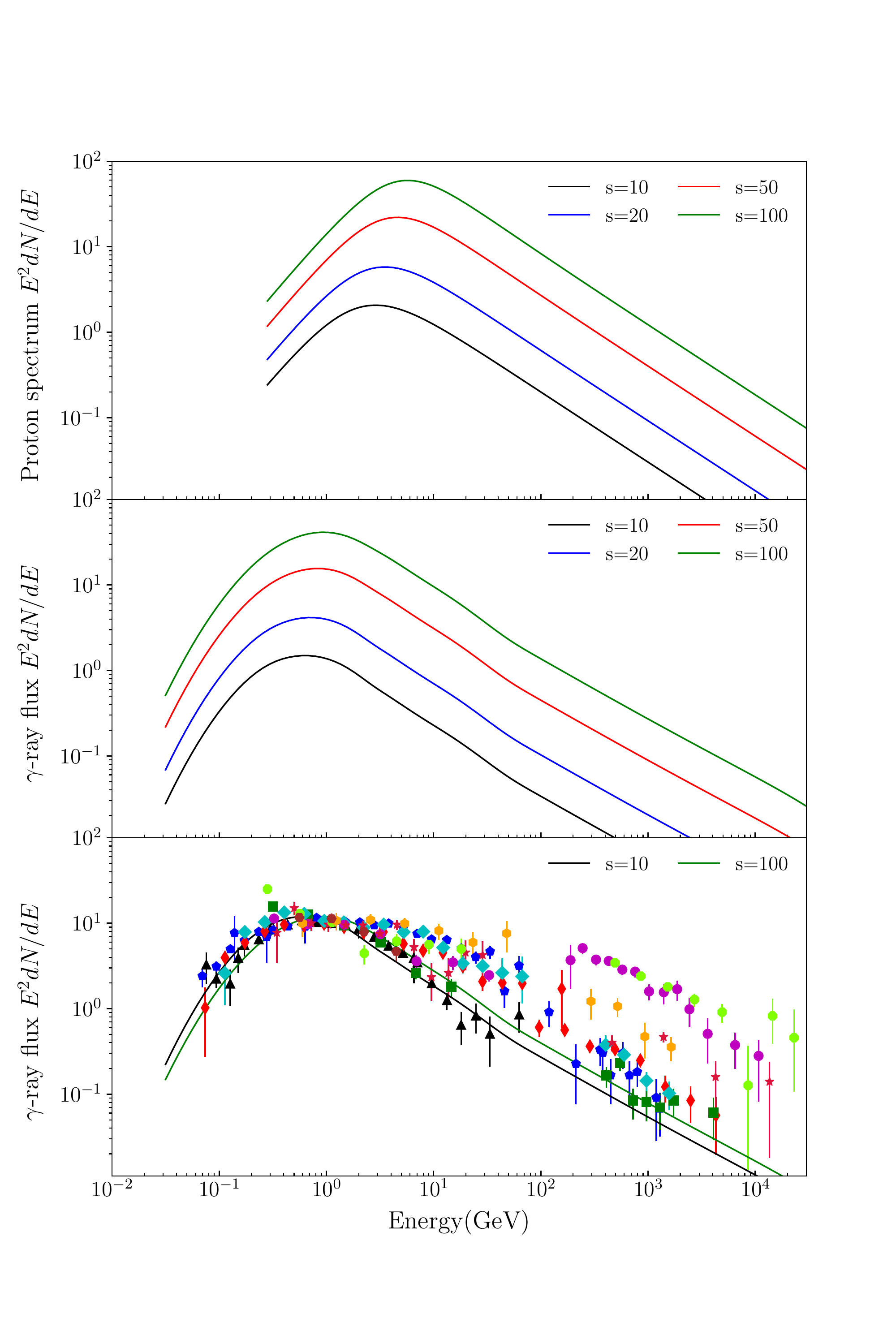} 
\caption{Same as Fig. \ref{fig:escape}, but the primary CR proton spectrum is taken from eq. (\ref{eq:ad_spectrum}) due to pure adiabatic compression of ambient CRs. $s$ is the adiabatic compression ratio.} 
    \label{fig:adiabatic}
\end{center}
\end{figure}

Assuming pure adiabatic compression, the diversity of SNR properties and ISM environment in SNR/MC is simply reflected in the compression ratio $s$. In Fig. \ref{fig:adiabatic}, we plot the primary CR proton spectrum based on eq. (\ref{eq:ad_spectrum}) (upper panel) and the corresponding $\pi^0$-decay emission (middle panel) for different $s$. When $s$ increases from 10 to 100, the compressed CR spectrum is boosted significantly in the vertical y-axis, while the shift of spectrum in the horizontal x-axis is relatively small. In the lower panel of Fig. \ref{fig:adiabatic}, we present the scaled  $\gamma$-ray data and model spectra which are normalized to have the same peak value. It is shown that the spectral shape of $\pi^0$-decay emission based on pure adiabatic compression remains almost unchanged with different $s$.  The compressed CR spectrum is capable to explain the spectral shape of SNR/MC like W44 and IC 443, which are consistent with the conclusion in \cite{TC14} and \cite{Cardillo16}. But it has difficulty in reproducing the emission of SNR/MC with harder TeV spectrum.  It might imply that the ambient CR spectrum in the vicinity of SNRs is harder than the local values measured in our solar neighborhood, which will be discussed in section \ref{sec:PCR_spectrum}.


The pure adiabatic compression case might correspond to the situation with a quasi-perpendicular shock, where DSA is inefficient. Hybrid simulations show that at very oblique shocks ions only gain a factor of a few in momentum and energy through shock drift acceleration \citep{CS14}. Hence, in direct interaction scenario the $\gamma$-ray emission from SNR/MC like W44 is possibly explained by shock drift acceleration in quasi-perpendicular shock plus adiabatic compression. The main challenge for such an explanation is the origin of quasi-perpendicular shock. If the dense filaments in MCs have a preferred magnetic field direction along the filaments, then quasi-perpendicular shock might form during the interaction between SNR and the dense filaments.

\subsection{DSA of thermal injected particles}
In this section, we discuss about the situation involving DSA of thermal injected particles. In young SNRs, it is widely accepted that the seed particles injected into DSA process are those energetic particles in the high energy tail of thermal population \citep[e.g.][]{Blasi13}. In middle aged SNRs, as a first guess it is natural to adopt the same assumption. However, as we will show later, thermal injected particles as seed particles have difficulties in reproducing the observed TeV emission.

The CR spectrum resulting from DSA of thermal injected particles is assumed to be a steady state DSA spectrum \citep{Bell78,B&E87} plus a spectral break at $p_{br}$ and an exponential cutoff at $p_{max}$, which is
\begin{equation}
n_{DSA}(p)= \left\lbrace
\begin{aligned}
 N_0p^{2-\alpha_r}e^{-p/p_{max}},& &p\leq p_{br},\\
  N_0p_{br}p^{1-\alpha_r}e^{-p/p_{max}},& &p> p_{br}.
\end{aligned}\right.
\label{eq:DSA_TI}
\end{equation}
$N_0$ is a normalization constant and $\alpha_r=3r/(r-1)$ where $r$ is the shock compression ratio. Strong shock condition is assumed through our discussion, i.e. $r=4$. 

The spectral break at $p_{br}$ is introduced by a modified version of neutral-ion damping. \cite{Malkov11} propose that neutral-ion damping can steepen the steady state DSA spectrum by exactly one power above the break momentum $p_{br}$, which help explain the steeping of $\gamma$-ray emission above a few GeV. The break momentum $p_{br}$ is found to be \citep{Malkov11}
\begin{eqnarray}
p_{br}&=& \frac{2V_A m \omega_c}{\nu_{in}} \nonumber \\
&\approx & 9  ~B_{a,0}^2T^{-0.4}_4n_n^{-1}n_i^{-1/2} ~{\rm GeV/c},
\end{eqnarray}
where $T_4$ is the precursor temperature in $10^4$K, $B_{a,0}$ is the ambient magnetic field in $\mu G$, $n_n$ and $n_i$ are the number density of neutrals and ions in $\rm cm^{-3}$ respectively. 

The exponential cutoff at $p_{max}$ represents the maximum energy available in accelerated particles due to wave damping, energy loss and finite acceleration time.
The evolution of $p_{max}$ due to either non-linear wave damping or neutral-ion damping in a young SNR is discussed in section \ref{sec:GeVescape}. Here we assume $ p_{max}$ is limited by the finite acceleration time of particles like in \cite{Uchiyama10}, which could be considered as an upper limit of $p_{max}$. The condition that particle acceleration time is smaller than the remnant age provides
\begin{equation}
p_{max}\lesssim 500\,\eta_g^{-1}u_{sh,2}^2t_4B_{a,1}{ \rm GeV/c },
\label{eq:cutoff_p}
\end{equation}
where $\eta_g\geq 1$ is the gyro factor, $t_{4}$ is the remnant age in $10^4$yr and $B_{a,1}$ is the ambient magnetic field in $10\mu G$.

\begin{figure}
\begin{center}
\includegraphics[width=\columnwidth]{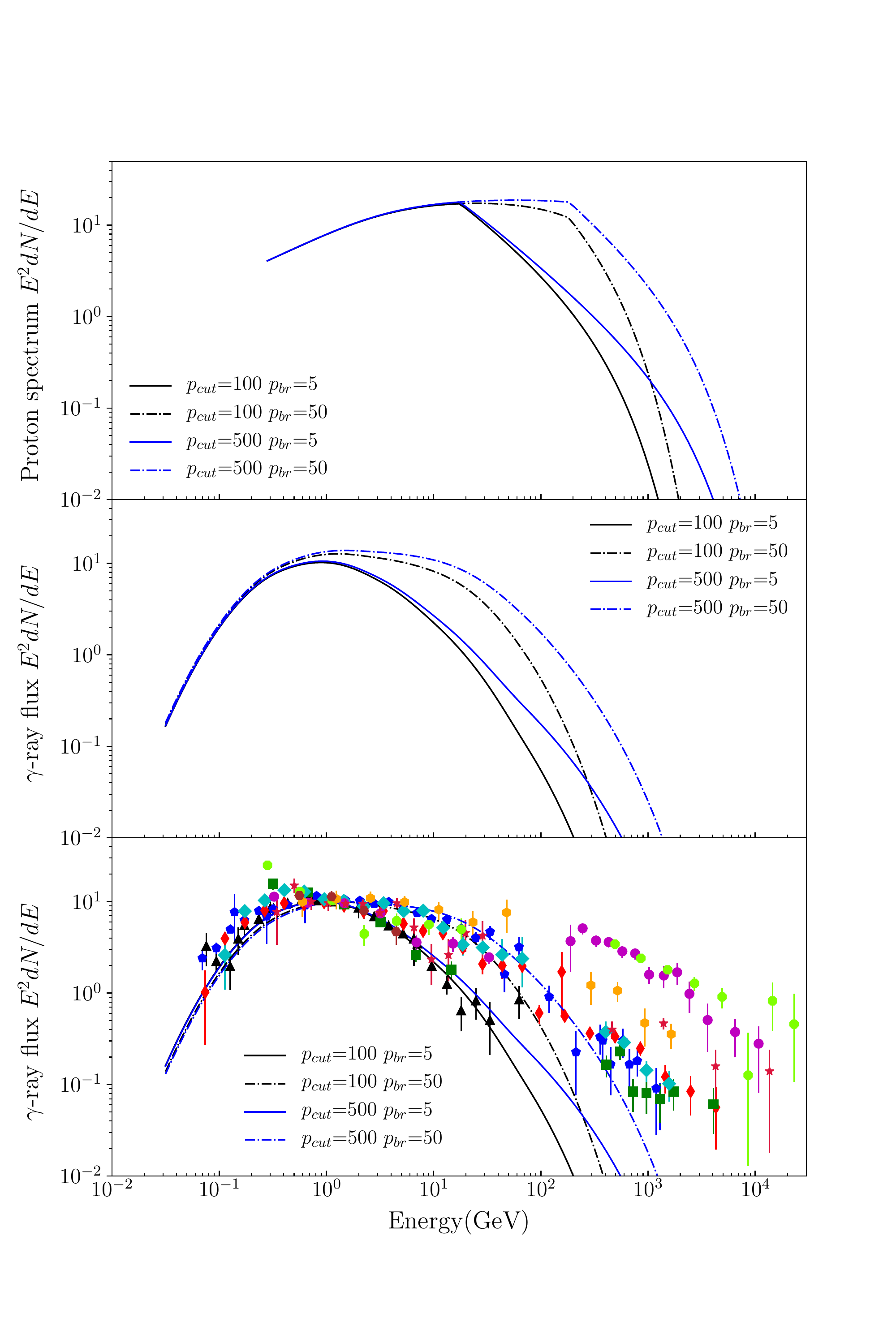} 
\caption{Same as Fig. \ref{fig:escape}, but the primary CR proton spectrum is described by  eq. (\ref{eq:DSA_TS}) based on DSA of thermal injected seed particles. $p_{max}$ and $p_{br}$ are the cutoff momentum and break momentum in GeV/c respectively. The adiabatic compression ratio $s$ is assumed to be 50.} 
    \label{fig:DSA_thermal}
\end{center}
\end{figure}

When these energetic CR particles accumulate in the radiative shell, the CR spectrum derived in eq. (\ref{eq:DSA_TI}) is further boosted by adiabatic compression and becomes
\begin{equation}
n_{ad,DSA}(p)=s^{2/3}n_{DSA}(s^{-1/3}p).
\label{eq:DSA_TS}
\end{equation}
In Fig. \ref{fig:DSA_thermal}, we plot the primary CR proton spectrum $n_{ad,DSA}$ (upper panel) and the corresponding $\pi^0$-decay emission (middle panel) for different $p_{max}$ and $p_{br}$. The scaled $\gamma$-ray data and model spectra are presented in the lower panel to compare the spectral shape. The adiabatic compression ratio $s$ is assumed to be 50 in all the calculation. 

According to Fig. \ref{fig:DSA_thermal}, the spectral break $p_{br}$ induced by neutral-ion damping is crucial in interpreting the steepening of $\gamma$-ray spectrum above a few GeV. The main concern about the modified version of neutral-ion damping is the time scale taken for it to rebuild the CR spectrum, which is not discussed in \cite{Malkov11}. As a result, whether such modified version of neutral-ion damping is efficient enough in middle aged SNRs is still an open question \citep{Drury11}. 

The $\gamma$-ray data from all SNR/MC suggests that $p_{br}\sim 10$GeV/c. If we assume $n_0=n_i+n_n$ and the ionization fraction is $\theta$, then the break momentum becomes
\begin{equation}
p_{br}\approx  9  ~B_{a,0}^2n_0^{-3/2}T^{-0.4}_4(1-\theta)^{-1}\theta^{-1/2}~{\rm GeV/c}. 
\end{equation}
Zeeman measurements of diffuse and molecular clouds show that the total magnetic field within clouds in $\mu G$ follows \citep{Crutcher12}
\begin{equation}
B_{a,0}\approx\left\lbrace \begin{array}{ll}
C_0 \qquad \qquad \qquad\mbox{when }n_0<300,\\
C_0(n_0/300)^{0.65} \quad \mbox{when } n_0\geq 300,\\
\end{array}
\right.
\end{equation}
where $n_0$ is the ambient density in $\rm cm^{-3}$, $C_0$ is a constant and $1\lesssim C_0\lesssim 10$.  
In dense MCs, we have
\begin{equation}
p_{br}\approx  0.2  ~C_0^2 n_0^{-0.2}T^{-0.4}_4(1-\theta)^{-1}\theta^{-1/2}~{\rm GeV/c}, 
\end{equation}
which is not very sensitive to $n_0$. Because $T_4$ and $\theta$ both depend on the shock velocity $u_{sh}$, $p_{br}\sim 10$GeV/c then implies a shock velocity $u_{sh}\sim 100$km/s and 
$C_0 \sim 5$ which can be tested by future observation.

The primary CR spectrum derived in eq. (\ref{eq:DSA_TS}) is able to reproduce the $\gamma$-ray emission in W44 and G357.7-0.1, which do not have TeV detection \citep{Uchiyama10}. For the rest of SNR/MC with TeV detection, the model discussed in this section however fails in the TeV band due to the exponential cutoff \citep{TC14}. It is mainly because DSA is inefficient in middle aged SNRs with slow shocks and is not able to accelerate thermal injected particles to above TeV energy, which suggests that thermal injected particles might not be the dominant component of seed particles in SNR/MC. In section \ref{sec:ad_compression}, we already show that pure adiabatic compression of ambient CRs is capable to produce a significant amount of TeV emission, which motivates us to consider ambient CRs as seed particles in DSA, i.e. re-acceleration of pre-existing ambient CRs.

\subsection{Re-acceleration of pre-existing ambient CRs}{\label{sec:RPCR}}
In this section, we discuss the model involving re-acceleration of pre-existing ambient CRs (hereafter RPCR). From energetic point of view, in middle-aged SNRs with slow shocks RPCR is more efficient in accelerating particles. Recently, \cite{Lee15} performed a hydro simulation for W44 with a self-consistent treatment of non-linear DSA. The authors found that, in order to reproduce the observed $\gamma$-ray and radio emission, about $\sim 33\%$ of shock energy has to contribute to the DSA process for thermal injection, while only $\lesssim 1\%$ of acceleration efficiency is needed for RPCR. 

If RPCR is the correct picture for DSA in middle-aged SNRs, then it indicates a transition of seed particles during the SNR evolution, which is from thermal injected seed particles in young SNRs to pre-existing CRs in middle-aged SNRs. The transition is a natural consequence of the declining shock velocity in the SNR evolution. For slow shock, the postshock temperature is low, thus the thermal particles behind the shock front are shifted to lower energy, which makes the thermal injection less efficient. 

RPCR is potentially very interesting as it may help explain the anomaly detected by PAMELA and AMS in the ambient CR spectrum \citep[e.g.,][]{Adriani11}. For example, \cite{TH14} propose that strong re-acceleration of ambient CRs at slow SNR shocks is able to explain the observed spectral break in CR proton and Helium spectra at a few hundred GV rigidity. The study here provides possibly the first evidence for efficient RPCR in middle-aged SNRs.

Next, we try to estimate the condition for such transition to happen based on simple energy argument. We assume that particles with $E>\xi E_{th}$ are injected into the DSA process as suggested by \cite{Blasi05}, where $E_{th}=k_BT$ characterizes the thermal peak of the Maxwellian distribution. Let's consider a shock front with velocity $u_{sh}$, which is propagating in an ambient medium with number density $n_a$. Under the strong shock condition, the postshock density and temperature are $n_{sh}=4n_a$ and $T_{sh}=3\mu m_{p}u^2_{sh}/16k_B$ respectively, where $m_p$ is the proton mass and $\mu$ is the mean molecular weight. The total energy density of thermal injected protons at the downstream region is
\begin{eqnarray}
E_{inj}&=&\frac{2n_{sh}}{\sqrt{\pi}}\int^\infty_{\xi E_{th}}\left( \frac{E}{k_B T_{sh}}\right)^{3/2}e^{-E/k_B T _{th}}dE \nonumber\\
&=&\frac{3\mu m_p n_{a}u^2_{sh}}{2\sqrt{\pi}}\int^\infty_{\xi}x^{3/2}e^{-x}dx \nonumber \\
&=&420 \mu\left(\frac{n_a}{1\rm cm^{-3}}\right)\left(\frac{u_{sh}}{200 \rm km/s}\right)^2 \rm eVcm^{-3}\nonumber\\
&&\times \int^\infty_{\xi}x^{3/2}e^{-x}dx.
\end{eqnarray}
\cite{Blasi05} argue that $\sqrt{\xi}$ varies between 2 and 4. Assuming $\sqrt{\xi}=3$ and $\mu=1.4$, after some calculation we obtain 
\begin{equation}
E_{inj}=2.3\left(\frac{n_0}{1\rm cm^{-3}}\right)\left(\frac{u_{sh}}{200 \rm km/s}\right)^2 \rm eVcm^{-3},
\label{eq:seed_transition}
\end{equation}
which is comparable to the ambient CR energy density $E_{CR}\sim 1\rm eVcm^{-3}$. We want to emphasize that $E_{inj}$ is very sensitive to $\xi$ and decreases very quickly as $\xi$ increases. If we assume $\sqrt{\xi}=4$, we instead obtain 
\begin{equation}
E_{inj}=0.005\left(\frac{n_0}{1\rm cm^{-3}}\right)\left(\frac{u_{sh}}{200 \rm km/s}\right)^2 \rm eVcm^{-3}.
\end{equation}
According to Eq. (\ref{eq:seed_transition}), ambient CRs likely become the dominant component of seed particles in DSA, if the shock velocity drops to $u_{sh}\sim 100$km/s. It is interesting that the slow radiative shock observed in middle-aged SNRs like IC 443 and W44 is consistent with the requirement of RPCR. Although above calculation may not be the best way to estimate the transition of seed particles, it qualitatively illustrates the existence of such transition. Quantitative study of the transition in a self consistent way is beyond the scope of this paper, which will be addressed in future work.

\cite{TC15} derive time dependent solutions for RPCR with both energy independent diffusion and energy dependent diffusion in the test particle limit. In this section, we confine our discussion to energy dependent diffusion which is more realistic in SNR/MC. In \cite{TC15}, we adopt the spatially averaged CR spectrum in the downstream region to fit the $\gamma$-ray emission. Here we instead apply the particle spectrum at shock front
to fit the $\gamma$-ray spectra, which might be more reasonable for a thin shell. The RPCR spectrum at shock front is 
\begin{align}
n_{TD}(p)
&=\alpha p^{2-\alpha_r} \int_0^p p'^{\alpha_r-3} n_{CR}(p')dp' \nonumber\\
&\times\mathcal{L}^{-1}\left\lbrace \frac{e^{2\Delta \alpha(\sqrt{1+sp'_\sigma }-\sqrt{1+sp_\sigma})/A_2\sigma} }{2s} \right.  \nonumber \\
&\times \left. \frac{(1+\sqrt{1+sp_\sigma})^{(2\Delta \alpha / A_2\sigma)}}{(1+\sqrt{1+sp'_\sigma})^{(2\Delta \alpha / A_2\sigma)-1}}\right\rbrace,
\label{eq:TD_spectrum}
\end{align}
where $\mathcal{L}^{-1}$ represents inverse Laplace transformation, $n_{CR}(p)$ is the ambient CR spectrum defined in eq. (\ref{eq:CR_spectrum}) and $\sigma$ is the power law index for energy dependent diffusion. The definition of all the parameters in above equation can be found in \cite{TC15}. The RPCR spectrum provided in eq. (\ref{eq:TD_spectrum}) is characterized by a critical momentum $p_{crit}$, which is determined by the acceleration time of CR particles and is equivalent to $p_{max}$ defined in eq. (\ref{eq:cutoff_p}). Below $p_{crit}$, the particle spectrum already reaches the steady state and follows the steady state DSA solution. Above $p_{crit}$, the particle spectrum hasn't reached the steady state yet and instead follows the input CR spectrum, see \cite{TC15,TC16} for more details. 

\begin{figure}
\begin{center}
\includegraphics[width=\columnwidth]{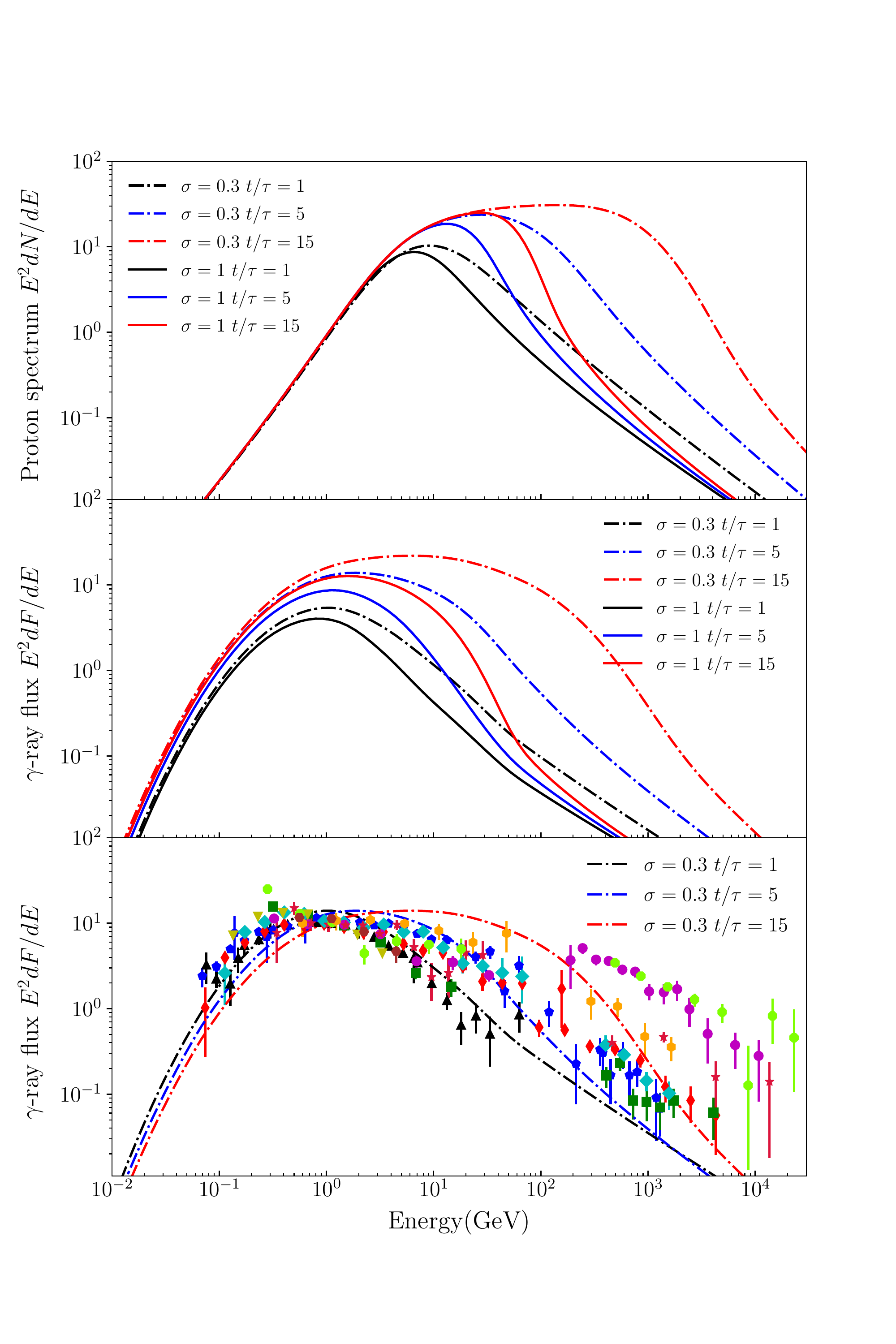} 
\caption{Same as Fig. \ref{fig:escape}, but the primary CR proton spectrum is replaced with  eq. (\ref{eq:ad_TD}) involving both RPCR and adiabatic compression. $\sigma$ is the power law index for energy dependent diffusion. $t/\tau$ is a dimensionless time ratio, where $t$ is roughly the remnant age and $\tau$ is a characteristic time for DSA, see eq. (\ref{eq:characteristic_time}). The adiabatic compression ratio $s$ is assumed to be 50.} 
    \label{fig:RPCR}
\end{center}
\end{figure}

In the radiative shell, CR particles generated through RPCR are further boosted by adiabatic compression. The resulted primary CR spectrum involving both RPCR and adiabatic compression becomes
\begin{equation}
n_{ad,TD}(p)=s^{2/3}n_{TD}(s^{-1/3}p).
\label{eq:ad_TD}
\end{equation}
In Fig. \ref{fig:RPCR}, we plot the primary CR spectrum $n_{ad,TD}$ (upper panel) and the corresponding $\pi^0$-decay emission (middle panel) for different $\sigma$ and $t/\tau$. $t/\tau$ is a dimensionless ratio between the particle acceleration time $t$ and the characteristic time $\tau$ for DSA, which satisfies
\begin{equation}
\frac{t}{\tau}\sim 0.3 \,\left( \frac{t}{10^4 \rm yrs}\right)\left(\frac{u_{sh}}{100 \rm km/s} \right)^2 \left(\frac{10^{25}\rm cm^2/s}{D_0}\right).
\label{eq:characteristic_time}
\end{equation}
$D_0$ is diffusion coefficient at $p=1{\rm GeV/c}$ and $u_{sh}$ is shock velocity \citep{TC15}.  
In the lower panel, we present the $\gamma$-ray data and model spectra, which are scaled to have the same peak value. The adiabatic compression ratio $s$ is assumed to be 50 in all cases. 

According to Fig. \ref{fig:RPCR}, as the ratio $t/\tau$ increases the critical momentum $p_{crit}$ in the proton spectrum is shifted to higher energy, while the power law index below and above the break remain almost the same, see more discussion in \cite{TC15}. This feature in RPCR can naturally explain why the BPL spectrum in SNR/MC has similar $\alpha_1$ and $\alpha_2$ but larger $p_{br}$ comparing with that in isolated giant MCs as discussed in section \ref{sec:spectral_property}.

In the case of RPCR, the modified version of neutral-ion damping \citep{Malkov11} is not needed to explain the steepening of spectrum above a few GeV \citep{TC15,Cardillo16}. According to Fig. \ref{fig:RPCR}, the model involving both RPCR and adiabatic compression is able to reproduce the overall profile of $\gamma$-ray spectra in all SNR/MC except W30 and W41. The dispersion shown in the TeV spectra is likely partly explained by the different acceleration time, i.e. $t/\tau$, and partly induced by the intrinsic dispersion in the ambient CR spectrum, which will be discussed in the following section. W30 and W41 show much harder TeV emission than the others which is possibly caused by the spatially overlapped PWN in the line of sight.

The particle spectrum presented in eq. (\ref{eq:TD_spectrum}) can be approximated as 
\begin{equation}
n_{ap}(p)=\mbox{Max}[n_{CR}(p), n_{acc}(p)],
\label{eq:RPCP_spectrum}
\end{equation}
where $n_{CR}(p)$ is the ambient CR spectrum. $n_{acc}$ is the steady state DSA spectrum \citep{Bell78,B&E87} with an exponential cutoff at $p_{max}$
\begin{equation}
n_{acc}(p)=\alpha_r p^{2-\alpha_r}e^{-p/p_{max}}\int^p_{p_{min}}n_{CR}(p')p'^{\alpha_r-3}dp', 
\end{equation}
where $\alpha_r=3r/(r-1)$ and $r$ is the shock compression ratio. $p_{max}$ here is the same as that in eq. (\ref{eq:cutoff_p}) and is equivalent to $p_{crit}$. $n_{ap}(p)$ follows the steady state DSA spectrum below $p_{max}$ and approaches the input CR spectrum above $p_{max}$. $p_{min}$ is the minimum momentum in the ambient CR spectrum, which is negligible for our discussion. \cite{Lee15} and \cite{Cardillo16} found that the approximate CR spectrum  provided in eq. (\ref{eq:RPCP_spectrum}) plus adiabatic compression is able to reproduce both the radio and $\gamma$-ray emission in W44. 


\subsection{Pre-existing ambient CR spectrum}\label{sec:PCR_spectrum}
For SNR/MC with hard TeV spectrum, the ambient CR spectrum measured in our solar neighborhood seems to be too steep to explain the observed TeV emission. It might imply that the ambient CR spectrum in the vicinity of a SNR is harder than the local value. There are several possibilities which can induce a harder ambient CR spectrum. 

At first, the ambient CRs surrounding a SNR is strongly affected by the escaping CR particles, which can potentially harden the ambient CR spectrum. As particles with higher energy escape from the remnant more easily. A self-consistent model involving both ambient CRs and escaping CR particles is needed in future to investigate this possibility. Next, the CR background in our Galaxy may not be uniform as assumed in the standard model of Galactic CRs. It is possibly linked to the very high energy $\gamma$-ray emission discovered by H.E.S.S in the Galactic Centre ridge \citep{Aharonian06} and the excess of GeV emission discovered by Fermi in the inner Galaxy \citep{Ackermann12}. One possible explanation for the excess of diffuse $\gamma$-ray emission is that the CR background in our Galaxy is not uniform and the ambient CR spectrum in the Galactic central region is harder than our local value. 
If the ambient CR spectrum is not uniform in our Galaxy, then the dispersion shown in TeV emission of all SNR/MC simply reflects the non-homogeneity of CR background in our Galaxy. Ambient CR spectrum with spectral index of $2.5-2.7$ at very high energy would be enough to explain the TeV emission from different SNR/MC.

\subsection{Challenge for the model}\label{sec:challenge2}
In the direct interaction scenario, RPCR plus adiabatic compression appears to the best model to explain the $\gamma$-ray emission in SNR/MC. RPCR suggests a transition of seed particles in SNR evolution, which is from thermal injected seed particles in young SNR to the pre-existing ambient CRs in middle-aged SNR. In the transition, the spectrum of SNR/MC gradually changes from an exponential like cutoff at very high energy to a power law profile as the remnant age increases. 

The steep GeV spectrum and non-detection of TeV emission in W44 imply that it may be a good example of SNRs in such transition. It is likely that the thermal injected particles in W44 still dominate the seed particles, while ambient CRs already become the dominant component of seed particle in the other middle aged SNRs. Multi-wavelength observation of both young and middle-aged SNRs is needed in future to test the idea of transition in seed particle.

\section{Discussion}\label{sec:discussion}
In this work, we study the $\gamma$-ray emission from 10 SNR/MC in the First {\it Fermi} SNR catalog with a focus on the spectral shape. We compare the $\gamma$-ray data available in literature with the model prediction from widely used escaping scenario and direct interaction to obtain deeper insight into the physical origin of $\gamma$-ray emission. 

In the escaping scenario, the shape of model spectrum strongly depends on the spatial configuration of SNR/MC. Since the observed 10 SNR/MC are likely in different spatial configuration, a variety of $\gamma$-ray spectra with different shape is expected in observation, which however is not seen in current data. Moreover, in order to reproduce the $\gamma$-ray data, CR protons with energy $\lesssim 1$GeV must be able to escape from the remnant and then diffuse into the nearby MCs, which however is still an open question. Two best examples of illuminated MCs discussed in the literature are $\gamma$-ray bright MCs external to W28 and W44. We argue that the $\gamma$-ray emission from them is possibly attributed to or at least partly attributed to the ambient CRs and other associated sources. In summary, the $\gamma$-ray spectra from 10 SNR/MC are inconsistent with the prediction of escaping scenario statistically. The inconsistency not only challenges the escaping scenario but also challenges our understanding of CR escaping in SNRs. It may imply that the free escape boundary widely adopted in the escaping scenario is not a good recipe to describe the spatial distribution of escaping CRs. We have to keep in mind that the free escape boundary was originally introduced in non-linear DSA to achieve a self-consistent treatment of particle acceleration at the shock front. It is not designed to study the propagation of escaping CRs in the ISM surrounding a SNR. If we instead assume a finite escape probability at the forward shock for particle with all energies, then the sharp low energy cutoff in the escaping CR spectrum will disappear, which may relieve the tension between $\gamma$-ray data and the escaping model.

In the direct interaction scenario, the model involving RPCR and adiabatic compression is able to explain the $\gamma$-ray emission from most SNR/MC. RPCR suggests a transition in seed particle, which is from thermal injected seed particle in young SNRs to pre-existing ambient CRs in middle-aged SNRs. The transition is likely a natural result of declining shock velocity in SNR evolution. Because in old SNRs with slow shock the thermal particles behind the shock front are shifted to lower energy, which makes the thermal injection less efficient. Re-acceleration of ambient CRs is potentially very interesting as it may help explain the anomaly detected by PAMELA and AMS in the ambient CR spectrum \citep[e.g.,][]{Adriani11}. For example, \cite{TH14} propose that re-acceleration of ambient CRs at slow SNR shocks is able to explain the observed spectral break in proton and Helium spectra at a few hundred GV rigidity. Multi-wavelength observation is needed in future to investigate whether such transition exists in SNR evolution. We also propose that radiative SNRs without MC interaction are able to produce a significant amount of $\gamma$-ray emission. A good candidate is S147. With accumulated Fermi data and CTA in future, we expect to detect more remnants like S147. Through the discussion, we assume strong shock condition. If the slow shock in middle-aged SNRs happens to have a low Mach number, then it can reduce the compression ratio and steepen the steady state DSA spectrum, which needs to be investigated in future work.

\section*{Acknowledgments}
XT is very grateful to Roger Chevalier, Patrick Slane, Andrew Strong, Shigehiro Nagataki, Shiu-Hang Lee and the anonymous referee for constructive comments, which help to improve the manuscript significantly. XT would also like to thank Yang Chen, Xiao Zhang, Yasunobu Uchiyama and Eugene Churazov for reading the manuscript.

\clearpage

\appendix
\section{$\pi^0$-decay emisison}{\label{app:pion_decay}}
The $\pi^0$-decay emission from proton-proton interaction is calculated with the parameterized $\gamma$-ray production cross sections $d\sigma/dE_\gamma$ developed in \cite{Kafexhiu14}. At low energy, the model is fitted with experimental data while at high energy it is tested with public available code GEANT4, PYTHIA 8, SIBYLL 2.1 and QGSJET-I. The 4 public codes predict slightly different results at high energy. In this work, we apply the formula of $d\sigma/dE_\gamma$ fitted with GEANT 4 results to do the calculation. The analytical formula is found to be accurate within $20\%$ accuracy from the kinematic threshold ($280$MeV) to PeV energies. We notice that the $\gamma$-ray production cross sections provided by \cite{Kafexhiu14} are not smooth at some connecting points. But the resulted $\pi^0$-decay emission seems to be unaffected.  Emission from the secondary electrons are neglected for simplification in this paper.

The $\gamma$-ray production rate is
\begin{equation}
\frac{dF(E_\gamma)}{dE_\gamma}=4\pi n_a\int \frac{d\sigma}{dE_\gamma}(E, E_\gamma)J(E)dE,
\end{equation}
where $E_\gamma$ is the photon energy, $n_a$ is the number density of target protons, $E$ is the proton kinetic energy and $J(E)$ is the flux intensity of primary CR protons. $J(E)=vn(E)/4\pi$, where $v$ is proton velocity and $n(E)$ is the number density of primary CR protons. Note $E$ defined here is equivalent to $T_p$ in \cite{Kafexhiu14}.  

The parameterized $\gamma$-ray production cross section
\begin{equation}
\frac{d\sigma}{dE_\gamma}(E,E_\gamma)=A_{max}(E)\times F(E,E_\gamma)
\end{equation}
The analytical expressions of $A_{max}(E)$ and $F(E,E_\gamma)$ derived in \cite{Kafexhiu14} are quite complicated, so we only explain them briefly here. 

$A_{max}$ characterizes the maximum value of $d\sigma/dE_\gamma$ and depends on only the proton kinetic energy $E$. \cite{Kafexhiu14} found that
\begin{equation}
A_{max}=\left\lbrace\begin{aligned}
&b_0\frac{\sigma_\pi (E)}{E_\pi^{max}} & \mbox{if }E^{th}\leq E < 1\rm GeV \nonumber \\
&\frac{b_1m_p^{b_2-1}\sigma_\pi(E)}{E^{b_2}} e^{b_3\rm log^2(E/m_p)}	&\mbox{if } E\geq 1 \rm GeV\\
\end{aligned}\right.
\end{equation}
where $b_0=5.9$, $m_p$ is the proton rest energy, $E^{th}\approx 0.28$GeV is the threshold kinetic energy. $E_\pi^{max}$ is defined in eq. (10) and $b_1$-$b_3$ for different $E$ are presented in Table VII of \cite{Kafexhiu14}. $\sigma_\pi (E)$ is the inclusive $\pi^0$ production cross section and satisfies
\begin{equation}
\sigma_\pi (E)=\left\lbrace\begin{aligned}
&\sigma_{1\pi}(E)+\sigma_{2\pi}(E) & \mbox{if }E^{th}\leq E < 2\rm GeV \nonumber, \\
& \sigma_{inel}(E)\times \left\langle n_{\pi^0}\right\rangle (E) &\mbox{if } E\geq 2 \rm GeV.
\end{aligned}\right.
\end{equation}

$\sigma_{1\pi}$ is the cross section for $pp\rightarrow pp\pi^0$ channel and 
\begin{equation}
\sigma_{1\pi}=7.66\times 10^{-3}\,\eta^{1.95}(1+\eta+\eta^5)[f_{BW}(\sqrt{s})]^{1.86}{\rm mb}
\end{equation}
where $\eta $ and $f_{BW}(\sqrt{s})$ are provided in eq. (3) and eq. (4) of \cite{Kafexhiu14} respectively.

$\sigma_{2\pi}$ is the cross section for two-pion production channel and 
\begin{equation}
\sigma_{2\pi}=5.7 \left(1+e^{-9.3(E_1-1.4)}\right)^{-1}{\rm mb}
\end{equation}
where $E_1$ is proton kinetic energy in GeV.

$\sigma_{inel}$ is the total inelastic cross section for proton and proton interaction and 
\begin{align}
\sigma_{inel}=&\left[30.7 - 0.96{\rm log} \left( \frac{E}{E^{th}}\right) +0.18 {\rm log^2}\left(\frac{E}{E^{th}}\right)\right] \nonumber \\
&\times \left[1-\left(\frac{E^{th}}{E}\right)^{1.9} \right]^3 \rm mb.
\end{align}

$\left\langle n_{\pi^0}\right\rangle (E)$ is the average $\pi^0$ production multiplicity and
\begin{equation}
\left\langle n_{\pi^0}\right\rangle=
\left\lbrace\begin{aligned}
&-0.006+0.237Q_p -0.023Q_p^2, 
 &\mbox{if }2\leq E_1 < 5 \nonumber \\
&a_1\xi^{a_4}[1+e^{-a_2\xi^{a_5}}][1-e^{-a_3\xi^{1/4}}],
& \mbox{if } E_1\geq 5
\end{aligned}\right.
\end{equation}
where $Q_p=(E-E^{th})/m_p$, $\xi=(E-3GeV)/m_p$ and $E_1$ is proton kinetic energy in GeV. $a_1$ to $a_5$ are presented in Table IV of \cite{Kafexhiu14}.

$F(E,E_\gamma)$ describes the shape of $\pi^0$-decay spectrum and is a function of both $E$ and $E_\gamma$. \cite{Kafexhiu14} found that 
\begin{equation}
F(E,E_\gamma)=\left(1-X_\gamma^{\alpha(E)}\right)^{\beta(E)}\left(1+\frac{X_\gamma Y_\gamma^{max}}{\lambda(E) m_\pi}\right)^{-\gamma(E)}
\end{equation}
where $m_\pi$ is the rest energy of $\pi^0$ and 
\begin{equation}
X_\gamma=\frac{Y_\gamma-m_\pi}{Y_\gamma^{max}-m_\pi}.
\end{equation}
$Y_\gamma$ and $Y_\gamma^{max}$ are defined in eq. (9) and (10) of \cite{Kafexhiu14}, while $\lambda(E)$, $\alpha(E)$, $\beta(E)$ and $\gamma(E)$ for different $E$ are presented in Table V of \cite{Kafexhiu14}.

\section{Runaway CR spectrum}{\label{app:runaway_spectrum}}
The run away CR spectrum $f(R,t, p)$ at a given distance $R$ from the remnant center and at a given time $t$ since supernova explosion depends on several physical processes including SNR evolution, DSA and the recipe for CR escaping, which is assumed to be free escape boundary here. The resulted $\gamma$-ray emission due to interaction between run away CRs and particles in MCs further depends on the CR propagation in ISM and spatial distribution of MCs, i.e. shape and density profile of MCs. 

To simplify the problem, it is usually assumed that the whole system is in spherical symmetry, i.e. the SNR expands spherically with radius $R_{sh}$ and the nearby MCs spread in a spherical shell with inner radius $L_1$ and outer radius $L_2$.  It is also often assumed that the SNRs are evolving in the Sedov-Taylor phase and the MCs are uniform in density. According to free escape boundary condition, 
the time $t_{esc}$ for CR particles with momentum $p$ to escape the remnant has a power law dependence on $p$ and satisfies \citep[e.g.,][]{Ohira11}
\begin{equation}
t_{esc}(p)=t_{sedov}\left( \frac{p}{p_{knee}}\right)^{-5/2\alpha}.
\end{equation}
$p_{knee}$ is the momentum corresponding to CR knee energy, i.e., $p_{knee}\sim 3\times 10^{15}$eV/c, $t_{sedov}$ is the beginning time of Sedov-Taylor phase and $\alpha$ is a constant.
At $t=t_{esc}(p)$, CR particles with momentum $p$ are released at the free escape boundary with radius
$R_{esc}=(1+\kappa)R_{sh}$,
where $R_{sh}$ is the remnant radius at $t_{esc}$ and $\kappa$ is a constant. Assuming the remnant radius at $t_{sedov}$ is $R_{sedov}$, we can obtain
\begin{equation}
R_{esc}(p)=(1+\kappa)R_{sedov}\left( \frac{p}{p_{knee}}\right)^{-1/\alpha}.
\end{equation}

After escaping the remnant, the CR particles diffuse in the ISM with coefficient $D_{ISM}$. We assume $D_{ISM}$ for CR particles with momentum $p$ satisfies
\begin{equation}
D_{ISM}(p)=10^{28}\chi \left(\frac{cp}{10GeV} \right)^{\delta}\rm cm^2s^{-1}, 
\label{app:D_ISM}
\end{equation}
where $\chi$ and $\delta$ are constants. In our calculation, we assume $\chi = 1$ and $\delta =0.5$ which is close to the Galactic average value \citep{Berezinskii90}. By solving the diffusion equation of CR particles under free escape boundary condition, \cite{Ohira11} show that the spatial averaged CR spectrum from a spherical shell between $L_1$ and $L_2$ is
\begin{align}
N(p, t, L_1, L_2)=&\frac{3N_{esc}(p)}{8\pi (L_2^3-L_1^3)}\times \nonumber\\
&\left\lbrace\frac{1}{\sqrt{\pi}C_{esc}}\left[e^{-(C_1-C_{esc})^2}-
e^{-(C_2-C_{esc})^2} \right.\right. \nonumber \\
 &\left.\left. - e^{-(C_1+C_{esc})^2}+e^{-(C_2+C_{esc})^2} \right] \right.\nonumber\\
 &\left.  +{\rm erf}\left(C_2-C_{esc}\right) -{\rm erf}\left(C_1-C_{esc}\right)\right. \nonumber \\
&\left.   +{\rm erf}\left(C_2+C_{esc}\right)-{\rm erf}\left(C_1+C_{esc}\right) \vphantom{\frac{1}{2}}\right\}, 
\label{app:eq:spectrum}
\end{align}
where $C_{esc}=R_{esc}/R_d$, $ C_1= L_1/R_d$, $C_2=L_2/R_d$ and
${\rm erf}(x)=(2/\sqrt{\pi})\int^x_0e^{-z^2}dz $ is the error function. 
\begin{equation}
R_d=\sqrt{4D_{ISM}(t-t_{esc}(p))}
\end{equation}
characterizes the length scale that a particle with momentum $p$ travels ever since the escape.

$N_{esc}$ is the time integrated spectrum of escaping CRs and has a power law form
\begin{equation}
N_{esc}=A_{esc}p^{-w}.
\end{equation}
$w$ is a constant determined by DSA processes and $w=2.38$ is adopted here as in \cite{Ohira11}. $A_{esc}$ is a normalization constant and is left to be a free parameter in our calculation since we are only interested at the shape of run away CR spectrum.

In the upper panel of Fig. \ref{fig:escape}, we present  $N(E, t, L_1, L_2)$ as a function of proton kinetic energy $E$ for different remnant age $t$ and different distance $L_1$ of MCs. Through the paper, we fix the thickness of MCs to be $5$pc, i.e. $L_2-L_1=5$pc. In the calculation, we assume $R_{sedov}=2.1$pc and $t_{sedov}=210$yr as in \cite{Ohira11} for simplification. $N(E, t, L_1, L_2)$ and $N(p, t, L_1, L_2)$ are related by $N(E, t, L_1, L_2)=N(p, t, L_1, L_2)/\beta c$, where $\beta$ is proton velocity in $c$.
 
\bsp	
\label{lastpage}
\end{document}